\documentclass[twocolumn]{aastex62}
\usepackage{amsmath}
\hypersetup{
	colorlinks	= true,
	linkcolor	= red,
	urlcolor	= cyan,
	citecolor	= blue
}
\usepackage{amssymb}
\usepackage{lipsum}
\usepackage{multirow}

\usepackage[version=4]{mhchem}
\usepackage{ragged2e}
\usepackage{gensymb}
\usepackage{tikz}
\usepackage{graphics}
\usepackage[figure,figure*]{hypcap}
\usepackage[para,online,flushleft]{threeparttable}
\usepackage{comment}
\usepackage{tablefootnote}
\usepackage{subscript}

\makeatletter
\newcounter{reaction}
\renewcommand\thereaction{R\arabic{reaction}}
\@addtoreset{reaction}{chapter}
\newcommand\reactiontag%
{\refstepcounter{reaction}\tag{\thereaction}}
\newcommand\reaction@[2][]%
{\begin{equation}\ce{#2}%
\ifx\@empty#1\@empty\else\label{#1}\fi%
\reactiontag\end{equation}}
\newcommand\reaction@nonumber[1]%
{\begin{equation*}\ce{#1}\end{equation*}}
\newcommand\reaction%
{\@ifstar{\reaction@nonumber}{\reaction@}}
\makeatother

\graphicspath{{./images/}}

\turnoffediting

\shortauthors{Hu}

\begin{document}
	
	\title{Photochemistry and Spectral Characterization of Temperate and Gas-Rich Exoplanets}
	
	\correspondingauthor{Renyu Hu}
	\email{renyu.hu@jpl.nasa.gov \\ @2021 California Institute of Technology. \\ Government sponsorship acknowledged.}
	
	\author[0000-0003-2215-8485]{Renyu Hu}
	\affiliation{Jet Propulsion Laboratory, California Institute of Technology, Pasadena, CA 91109, USA}
	\affiliation{Division of Geological and Planetary Sciences, California Institute of Technology, Pasadena, CA 91125, USA}
	
	\begin{abstract}
Exoplanets that receive stellar irradiance of approximately Earth's or less have been discovered and many are suitable for spectral characterization. Here we focus on the temperate planets that have massive H$_2$-dominated atmospheres, and trace the chemical reactions and transport following the photodissociation of \ce{H2O}, \ce{CH4}, \ce{NH3}, and \ce{H2S}, with K2-18~b, PH2~b, and Kepler-167~e representing temperate/cold planets around M and G/K stars. We find that \ce{NH3} is likely depleted by photodissociation to the cloud deck on planets around G/K stars but remains intact in the middle atmosphere of planets around M stars. A common phenomenon on temperate planets is that the photodissociation of \ce{NH3} in presence of \ce{CH4} results in \ce{HCN} as the main photochemical product. The photodissociation of \ce{CH4} together with \ce{H2O} leads to \ce{CO} and \ce{CO2}, and the synthesis of hydrocarbon is suppressed. Temperate planets with super-solar atmospheric metallicity and appreciable internal heat may have additional \ce{CO} and \ce{CO2} from the interior and less \ce{NH3} and thus less \ce{HCN}. Our models of K2-18~b can explain the transmission spectrum measured by \textit{Hubble}, and indicate that future observations in $0.5-5.0\ \mu$m would provide the sensitivity to detect the equilibrium gases \ce{CH4}, \ce{H2O}, and \ce{NH3}, the photochemical gas \ce{HCN}, as well as \ce{CO2} in some cases. Temperate and \ce{H2}-rich exoplanets are thus laboratories of atmospheric chemistry that operate in regimes not found in the Solar System, and spectral characterization of these planets in transit or reflected starlight promises to greatly expand the types of molecules detected in exoplanet atmospheres.
	\end{abstract}
	
	
	\section{Introduction} \label{sec:intro}
	
The era of characterizing temperate exoplanets has begun. Kepler, K2, and TESS missions have found a few tens of exoplanets cold enough for water to condense in their atmospheres in transiting orbits (from the NASA Exoplanet Archive). Another handful of temperate planets may be confirmed in the next few years with ongoing validation and followup of TESS planet candidates \citep{barclay2018revised}. A small subset of these planets has been observed by HST for transmission spectra \citep{de2018atmospheric,zhang2018near,tsiaras2019water,benneke2019water}. For example, a transmission spectrum obtained by {\it Hubble} at 1.1 -- 1.7 $\mu$m of the temperate sub-Neptune K2-18~b shows spectral features \citep{tsiaras2019water,benneke2019water}, and the spectrum indicates that the planet hosts an atmosphere dominated by H$_2$, and has H$_2$O and/or CH$_4$ in its atmosphere \citep{benneke2019water,madhusudhan2020interior,Blain2020}. TOI-1231~b is another temperate planet suitable for atmospheric studies with transits \citep{Burt2021}. With $>7$ times more collecting area and infrared instruments, JWST will be capable of providing a more detailed look into the atmospheres of these temperate exoplanets \citep{beichman2014observations}.
	
We refer to the exoplanets that receive stellar irradiance of approximately Earth's as ``temperate exoplanets'' and those that receive less irradiance by approximately an order of magnitude as ``cold exoplanets'' in this paper. Temperate and cold exoplanets include both giant planets and small planets and potentially have diverse atmospheric composition. Giant planets (Jupiters and Neptunes) have massive H$_2$/He envelopes \citep[e.g.,][]{burrows2001theory}, and small planets (mini-Neptunes, super-Earths, and Earth-sized planets) can have H$_2$/He atmospheres with variable abundances of heavy elements, steam atmospheres mostly made of water, or secondary atmospheres from outgassing \citep[e.g.,][]{fortney2013framework,moses2013compositional,Hu2014B2014ApJ...784...63H}.
	
In this paper, we focus on temperate/cold and gas-rich exoplanets, which include temperate/cold giant planets and mini-Neptunes. We assume that the atmospheres are H$_2$/He-dominated and massive enough for thermochemical equilibrium to prevail at depths. This condition determines that the dominant O, C, N, S species should be H$_2$O, CH$_4$, NH$_3$, and H$_2$S on temperate and cold planets in most cases \citep[e.g.,][]{fegley1996atmospheric,burrows1999chemical,heng2016analytical,woitke2020coexistence}. Thermochemical equilibrium may also produce \ce{N2} as the dominant N species and substantial abundance of \ce{CO} and \ce{CO2} if the planet has a hot interior \citep[e.g.,][]{fortney2020beyond}. On temperate and cold planets, H$_2$O can condense to form a cloud and the above-cloud H$_2$O is partially depleted as a result \citep[e.g.,][]{morley2014water,hu2019information,charney2020}. Cold planets may additionally have NH$_4$SH (from the combination reaction between NH$_3$ and H$_2$S) and NH$_3$ condensed to form clouds \citep[e.g.,][]{lewis1969clouds,Atreya1999}. This paper primarily concerns the photochemical processes above the clouds, with H$_2$O, CH$_4$, NH$_3$, and H$_2$S as the feedstock.
	
Past work on the atmospheric photochemistry of low-temperature and gas-rich planets in the exoplanet context is rare. \cite{moses2016composition} studied the thermochemistry and photochemistry in directly imaged young giant planets, and discussed the photochemical production of \ce{CO2} and HCN in their atmospheres. \cite{zahnle2016photolytic} showed that sulfur haze can form photochemically in the young Jupiter 51 Eri b, and the level of the sulfur haze would moves upward in the atmosphere when the eddy diffusion coefficient increases. \cite{gao2017sulfur} further modeled the effect of the sulfur haze on the reflected starlight spectra of widely separated giant planets. Here we systematically study the atmospheric photochemistry of H$_2$O, CH$_4$, NH$_3$, and H$_2$S in low-temperature exoplanetary atmospheres and model the abundance of the photochemical gases to guide the future observations of temperate/cold and gas-rich exoplanets. 
	
The paper is organized as follows: Section~\ref{sec:model} describes the models used in this study; Section~\ref{sec:result} presents the results in terms of the main behaviors of atmospheric chemistry, key photochemical mechanisms, and the corresponding spectral features in transmission and reflected starlight; Section~\ref{sec:discussion} discusses the prospect to detect photochemical gases in temperate and gas-rich exoplanets and potential areas of further development; and Section~\ref{sec:conclusion} summarizes the key findings of this study.
	
\section{Methods} \label{sec:model}
	
\subsection{Atmospheric Structure Model} \label{sec:structure}
	
We use the atmospheric structure and cloud formation model in \cite{Hu2019B2019ApJ...887..166H} to simulate the pressure-temperature profile and potential gas depletion by condensation in temperate and cold exoplanets. 
	
We have updated the model with a routine to compute the condensation of NH$_4$SH cloud, in a similar way as the equilibrium cloud condensation model of \cite{atreya1985photochemistry}. In short, we compare the products of the partial pressure of NH$_3$ and H$_2$S with the equilibrium constant of the reaction that produces NH$_4$SH solid \citep{lewis1969clouds}, and partition the NH$_3$ and H$_2$S in excess to form the NH$_4$SH solid cloud in each atmospheric layer. We have verified that the resulting NH$_4$SH cloud density and pressure level is consistent with the previously published models when applied to a Jupiter-like planet \citep[e.g.,][]{Atreya1999}.
	
Another update is that the model now traces the concentration of NH$_3$ in liquid-water cloud droplets when applicable. The model of \cite{Hu2019B2019ApJ...887..166H} has included the dissolution of NH$_3$ in the liquid-water droplets. By additionally tracing the concentration of NH$_3$ in droplets, we have now taken into account the non-ideal effects when the NH$_3$ solution is non-dilute. When the mass ratio between NH$_3$ and H$_2$O in the droplet is $>0.05$, we replace Henry's law with the vapor pressure of NH$_3$ in equilibrium with the solution \citep{perry2007perry}. The latter merges the solubility in the Henry's law regime to that in the Raoult's law regime smoothly. We also apply the vapor pressure of H$_2$O in equilibrium with the solution, which can be substantially smaller than that with pure water when the solution is non-dilute (i.e., the Raoult's law). While the impact of these processes on the overall atmospheric composition of the planets studied in this paper -- planets warmer than Jupiter -- is small, these processes may control the mixing ratio of H$_2$O and NH$_3$ in the atmospheres of even colder planets \citep{de1989uranus,romani1989neptune}. 
	
\subsection{Atmospheric Photochemical Model}
\label{sec:photochemical}
	
We use the general-purpose photochemical model in \cite{hu2012photochemistry,hu2013photochemistry} to simulate the photochemical products in the middle atmospheres of temperate and cold exoplanets. The photochemical model includes a carbon chemistry network and a nitrogen chemistry network and their interactions \citep{hu2012photochemistry}. The photochemical model also includes a sulfur chemistry network and calculates the formation of \ce{H2SO4} and \ce{S8} aerosols when applicable \citep{hu2013photochemistry}.

We have made several updates to the original reaction network \citep{hu2012photochemistry}, and they are listed in Table~\ref{tab:reaction}. We have checked the main reactions that produce, remove, and exchange \ce{C1} and \ce{C2} hydrocarbons in the Jovian atmosphere \citep{gladstone1996hydrocarbon,moses2005photochemistry} and updated rate constants when more recent values in the relevant temperature range are available in the NIST Chemical Kinetics Database. We have added low-pressure or high-pressure rate constants for three-body reactions if any of them were missing in the original reaction rate list. Certain reactions important for the hydrocarbon chemistry do not have a directly usable rate constant expression in the NIST database; rather their rates are fitted on experimental data or estimated by \cite{moses2005photochemistry}. We have also added several reactions that involve \ce{NH} because it may be produced by \ce{NH3} photodissociation, and updated the rate constant of an important reaction \ce{NH2 + CH4 -> NH3 + CH3} to the latest calculated value. Lastly, we have removed two reactions that were incorrectly included: \ce{CH4 + C2H2 -> C2H3 + CH3} and \ce{C2H6 + C2H2 -> C2H3 + C2H5} because the reactant should have been \ce{CH2=C}.
	
The photochemical model is applied to the ``stratosphere'' of the atmosphere, where the ``tropopause'' is defined as the pressure level where the temperature profile becomes adiabatic. We define the lower boundary of the model as the pressure level 10-fold greater than the tropopause pressure, and thus include a section of the ``troposphere'' in the model. These choices are customary in photochemical studies of giant planets' atmospheres \citep[e.g.,][]{gladstone1996hydrocarbon}, and reasonable because the photochemical products in the stratosphere (and above the condensation clouds) are the objective of the study. Including a section of the troposphere makes sure that the results do not strongly depend on the lower boundary conditions assumed.
	
We apply fixed mixing ratios as the lower boundary conditions for \ce{H2}, \ce{He}, \ce{H2O}, \ce{CH4}, \ce{NH3}, and when applicable, \ce{H2S} according to assumed the elemental abundance. When interior sources of \ce{CO}, \ce{CO2}, and \ce{N2} are included in some scenarios (see Section~\ref{sec:planet} for detail), fixed mixing ratios are also applied to these gases at the lower boundary. We assume that all other species can move across the lower boundary (i.e., dry deposition when the lower boundary is a surface in terrestrial planet models) at a velocity of $K_{\rm zz}/H$, where $K_{\rm zz}$ is the eddy diffusion coefficient and $H$ is the scale height. This velocity is the upper limit of the true diffusion velocity, which could be damped by the gradient of the mixing ratio \citep{gladstone1996hydrocarbon}; however, the velocity only matters for long-lived species (e.g., \ce{C2H6} in Jupiter). Our choice of lower boundary conditions thus results in conservative estimates of the abundance of long-lived photochemical gases. 
	
The upper boundary is assumed at $10^{-4}$ Pa, i.e., small enough so that the peaks of photodissociation of all species are well within the modeled atmosphere. Following \cite{gladstone1996hydrocarbon}, we assume a zero-flux boundary condition for all species except for H, for which we include a downward flux of $4\times10^{9}$ cm$^{-2}$ s$^{-1}$ \citep{waite1983electron} to account for ionospheric processes that produce H. This influx of H was calculated for Jupiter and the actual flux can conceivably be different. The impact of this additional \ce{H} is limited to the upper atmosphere and, in most of our cases, is swamped by the \ce{H} from the photodissociation of \ce{H2O} (see Section \ref{sec:ch4}).
	
Since the modeled domain of the atmosphere includes the stratosphere and a small section of the upper troposphere, the standard mixing-length scaling \citep{gierasch1985energy} is not applicable to estimate the eddy diffusion coefficient. We instead anchor our choice of the eddy diffusion coefficient on the value in the upper troposphere of Jupiter ($\sim1\times10^{3}$ cm$^2$ s$^{-1}$, \cite{conrath1984global}) and explore a larger value in the study. Above the tropopause, we assume that mixing is predominantly caused by the breaking of gravity waves and the eddy diffusion coefficient is inversely proportional to the square root of the number density \citep{lindzen1981turbulence}.
	
Because the pressure range of the photochemical model typically includes the condensation of \ce{NH3} and \ce{H2O}, we have added a scheme to account for the condensation of \ce{NH3} into the photochemical model, with that for \ce{H2O} already included in the model of \cite{hu2012photochemistry}. In addition, we have added the schemes of condensation for \ce{N2H4} and \ce{HCN}, the two main photochemical gases expected to condense in Jupiter's upper troposphere \citep[e.g.,][]{atreya1977distribution,moses2010abundance}. The low-temperature vapor pressures of \ce{N2H4} and \ce{HCN} are adopted from \cite{atreya1977distribution} and \cite{krasnopolsky2009photochemical}, respectively. As such, these gases are treated in the photochemical model and their production and removal paths including chemical reactions and condensation are self-consistently computed. This is important because, for example, \ce{NH3} above the clouds in Jupiter is expected to be completely removed by photodissociation and converted to \ce{N2H4} and \ce{N2}, followed by condensation and transport to the deep atmosphere \citep{strobel1973photochemistry,atreya1977distribution,kaye1983formation,kaye1983hcn,moses2010abundance}. As we will show in Section~\ref{sec:result}, the condensation of \ce{N2H4} and \ce{HCN} limits their abundance in the middle atmosphere of cold planets like Kepler-167~e. For \ce{H2S}, we make a binary choice: if the cloud model indicates \ce{NH4SH} formation, we remove sulfur chemistry from the model, because \ce{NH4SH} should completely sequester \ce{H2S} \citep{atreya1985photochemistry}; and we include the sulfur chemistry if \ce{NH4SH} cloud is not formed. This simplifies the calculations of sulfur photochemistry and is broadly valid when N/S$>1$ in the bulk atmosphere.
	
We calculate the cross-sections and single scattering albedo of ammonia and water cloud particles using their optical properties \citep{palmer1974optical,martonchik1984optical} and the radiative properties of the sulfur haze particles in the same way as \cite{hu2013photochemistry}. \ce{NH4SH} and \ce{HCN} condensates are treated the same way as \ce{NH3} clouds. \ce{N2H4} condensates have very low abundance in all models and do not contribute significantly to the opacity. Thus, our model includes the absorption and scattering of cloud and haze particles when calculating the radiation field that drives photochemical reactions in the atmosphere.
	
	\startlongtable
	\begin{deluxetable*}{l|lll}
	    \tabletypesize{\scriptsize}
		\tablecaption{Reactions and rate constants updated with respect to \cite{hu2012photochemistry}. \label{tab:reaction}}
		\tablehead{
		\colhead{Reaction} & \colhead{Rate} & \colhead{Reason for Update} & \colhead{Source} }
		\startdata
		\ce{H + C2H3 -> C2H2 + H2} & $1.2\times10^{-12}T^{0.5}$ & Revise rate & \cite{moses2005photochemistry} \\
		\ce{H + C2H4 -> C2H3 + H2} & $5.0\times10^{-12}(T/298)^{1.93}\exp(-6520/T)$ & Revise rate & NIST \\
		\ce{H + C2H5 -> 2CH3} & $1.25\times10^{-10}$ & Revise rate & NIST \\
		\ce{CH + CH4 -> C2H4 + H} & $9.1\times10^{-11}(T/298)^{-0.9}$ ($T>295$ K) & Revise rate & NIST \\
		& $1.06\times10^{-10}(T/298)^{-1.04}\exp(-36/T)$ ($T\leq295$ K) & &  \\
		\ce{CH3 + H2 -> CH4 + H} & $2.31\times10^{-14}(T/298)^{2.24}\exp(-3220.0/T)$ & Revise rate & NIST \\
		\ce{C2H + CH4 -> C2H2 + CH3} & $1.2\times10^{-11}\exp(-490.7/T)$ & Revise rate & NIST \\
		\ce{C2H + C2H6 -> C2H2 + C2H5} & $2.58\times10^{-11}(T/298)^{0.54}\exp(180/T)$ & Revise rate & NIST \\
		\ce{C2H3 + H2 -> C2H4 + H} & $3.39\times10^{-14}(T/298)^{2.56}\exp(-2530.5/T)$ & Revise rate & NIST \\
		\ce{C2H3 + C2H5 -> 2C2H4} & $8.0\times10^{-13}$ & Revise rate & \cite{moses2005photochemistry} \\
		\ce{C2H3 + C2H5 -> C2H2 + C2H6} & $8.0\times10^{-13}$ & Revise rate & \cite{moses2005photochemistry} \\
		\ce{C2H5 + C2H5 -> C2H6 + C2H4} & $2.4\times10^{-12}$ & Revise rate & NIST \\
		\ce{NH + NH -> N2H2} & $8.47\times10^{-11}(T/298)^{-0.04}\exp(80.6/T)$ & Add reaction & NIST \\
		\ce{NH + NH2 -> N2H2 + H} & $1.5\times10^{-10}(T/298)^{-0.27}\exp(38.5/T)$ & Add reaction & NIST \\
		\ce{NH + CH4 -> NH2 + CH3} & $1.49\times10^{-10}\exp(-10103/T)$ & Add reaction & NIST \\
		\ce{NH + C2H6 -> NH2 + C2H5} & $1.16\times10^{-10}\exp(-8420.3/T)$ & Add reaction & NIST \\
		\ce{NH + OH -> NH2 + O} & $2.94\times10^{-12}(T/298)^{ 0.1}\exp(-5800/T)$ & Revise rate & NIST \\
		\ce{NH2 + CH4 -> NH3 + CH3} & $5.75\times10^{-11}\exp(-6952/T)$ & Revise rate & NIST \\
		\ce{H + H ->[M] H2}  & $k_0=\min(8.85\times10^{-33}(T/298)^{-0.6},1.0\times10^{-32})$ & Revise $k_0$; & NIST \\
		 & $k_{\infty}=1.0\times10^{-11}$ & Add $k_{\infty}$ & \cite{moses2005photochemistry}
		 \\
		 \ce{H + CH3 ->[M] CH4}  & $k_0=6.0\times10^{-29}\max(T/298,1.0)^{-1.8}$ & Revise $k_0$ at $T\leq298$ K; & NIST; \\
		 & $k_{\infty}=1.92\times10^{-8}(\max(T,110)^{-0.5}\exp(-400/\max(T,110))$ & Add $k_{\infty}$ & \cite{moses2005photochemistry}
		 \\
		 & $F_{c}=0.3+0.58\exp(-T/800)$ & & \\
		\ce{H + C2H ->[M] C2H2} & $k_0=1.26\times10^{-18}T^{-3.1}\exp(-721/T)$ & Add $k_0$ & \cite{moses2005photochemistry} \\
		& $k_{\infty}=3.0\times10^{-10}$ & & \\
		\ce{H + C2H2 ->[M] C2H3}  & $k_0=3.31\times10^{-30}\exp(-740/T)$ & Add $k_{\infty}$ & NIST \\
		 & $k_{\infty}=1.4\times10^{-11}\exp(-1300/T)$ & & \\
		 & $F_c=0.44$ & & \\
		\ce{H + C2H3 ->[M] C2H4} & $k_0=2.3\times10^{-24}T^{-1}$ & Add $k_0$; & \cite{moses2005photochemistry} \\
		& $k_{\infty}=1.8\times10^{-10}$ & Revise $k_{\infty}$ & \\
		\ce{H + C2H4 ->[M] C2H5}  & $k_0=\max(1.3\times10^{-29}\exp(-380/T),3.7\times10^{-30})$ & Revise $k_0$ at $T\leq300$ K; & NIST; \\
		 & $k_{\infty}=6.6\times10^{-15}T^{1.28}\exp(-650/T)$ & Add $k_{\infty}$ & \cite{moses2005photochemistry}
		 \\
		 & $F_{c}=0.24\exp(-T/40)+0.76\exp(-T/1025)$ & & \\
		 \ce{H + C2H5 ->[M] C2H6}  & $k_0=4.0\times10^{-19}T^{-3}\exp(-600/T)$ ($T>200$ K) & Revise $k_0$; & \cite{moses2005photochemistry} \\
		 & $k_{0}=2.49\times10^{-27}$ ($T\leq200$ K) &  &
		 \\
		 & $k_{\infty}=2.0\times10^{-10}$ &  &
		 \\
		\ce{C + H2 ->[M] CH2}  & $k_0=6.89\times10^{-32}$ & Add $k_{\infty}$ & NIST \\
		 & $k_{\infty}=2.06\times10^{-11}\exp(-57/T)$ & & \\
		 \ce{CH + H2 ->[M] CH3}  & $k_0=9.0\times10^{-31}\exp(550/T)$ & Revise $k_0$ and $k_{\infty}$ & NIST \\
		 & $k_{\infty}=2.01\times10^{-10}(T/298)^{0.15}$ & & \cite{moses2005photochemistry}
		 \\
		 \ce{CH3 + CH3 ->[M] C2H6}  & $k_0=1.68\times10^{-24}(T/298)^{-7}\exp(-1390/T)$ ($T>300$ K) & Revise $k_0$ at $T\leq300$ K; & NIST; \\
		 & $k_{0}=6.15\times10^{-18}T^{-3.5}$ ($T\leq300$ K) & Add $k_{\infty}$ & \cite{moses2005photochemistry}
		 \\
		 & $k_{\infty}=1.12\times10^{-9}T^{-0.5}\exp(-25/T)$ & & \\
		 & $F_{c}=0.62\exp(-T/1180)+0.38\exp(-T/73)$ & & \\
		\enddata
		\tablecomments{The rate constants of two-body reactions and the high-pressure limiting rate constants of three-body reactions ($k_{\infty}$) have a unit of cm$^3$ molecule$^{-1}$ s$^{-1}$, and the low-pressure limiting rate constants of three-body reactions ($k_0$) have a unit of cm$^6$ molecule$^{-2}$ s$^{-1}$. The rates of three-body reactions are $k_0k_{\infty}[{\rm M}]/(k_{\infty}+k_0[{\rm M}])F_c^\beta$, where $[{\rm M}]$ is the number density of the atmosphere, and $\beta=(1+(\log_{10}(k_0[{\rm M}/k_{\infty}))^2)^{-1}$. $F_c=0.6$ unless otherwise noted. NIST=NIST Chemical Kinetics Database (http://kinetics.nist.gov). }
	\end{deluxetable*}

\subsection{Jupiter as a Test Case}
\label{sec:jupiter}

	\begin{figure*}
	\centering
	\includegraphics[width=0.95\textwidth]{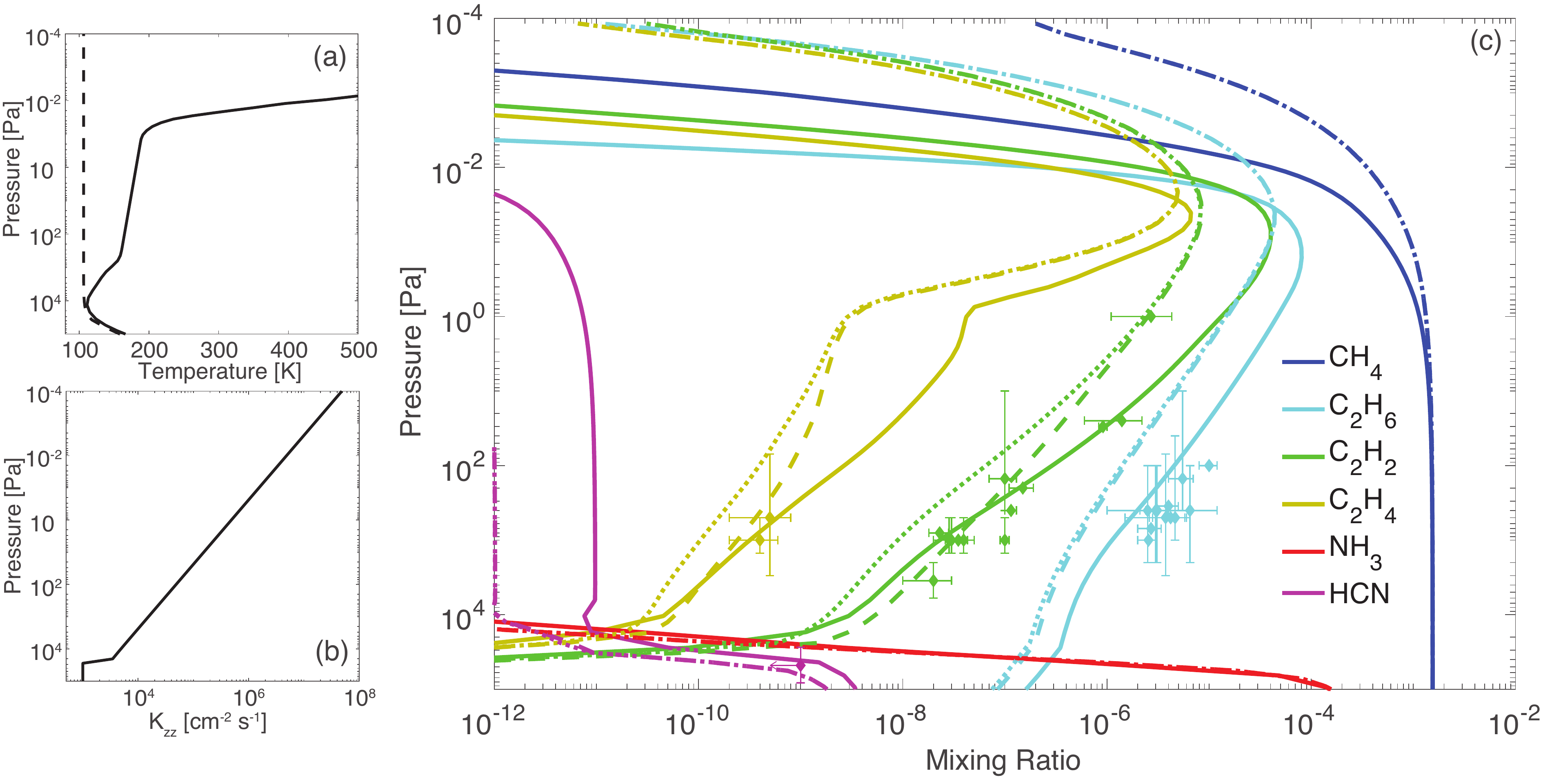}
	\caption{Jupiter as a test case. The planet modeled is a Jupiter-mass and Jupiter-radius planet at a 5.2-AU orbit of a Sun-like star, having an atmospheric metallicity of $3\times$solar. (a) The solid line is the pressure-temperature profile adopted from Galileo probe measurements and Cassini CIRS measurements in Jupiter \citep[the solid line;][]{seiff1998thermal,simon2006jupiter} and the dashed line is the pressure-temperature profile calculated by the atmospheric structure model. (b) The eddy diffusion coefficient profile adopted in this work. (c) The calculated mixing ratio profiles of \ce{CH4}, \ce{NH3}, and major photochemical products. The solid lines are the results using the measured temperature profile, the dashed lines are the results using the modeled temperature profile (i.e., without the temperature inversion), and the dotted lines are the results using the modeled temperature profile and the photodissociation quantum yield of \ce{C2H2} set to unity (see discussion in Section~\ref{sec:jupiter}). In comparison are the abundance data of major hydrocarbons and \ce{HCN} in Jupiter's atmosphere, as compiled in \cite{morrissey1995ultraviolet,gladstone1996hydrocarbon,davis1997broadband,yelle2001structure,moses2005photochemistry}.
	}
	\label{fig:jupiter}
	\end{figure*}
	
As a test case, we have applied the coupled cloud condensation and photochemical model to a Jupiter-like planet and compared the results with the measured gas abundance in Jupiter and previous models of Jupiter's stratospheric composition \citep{gladstone1996hydrocarbon,moses2005photochemistry,atreya1977distribution,kaye1983formation,kaye1983hcn,moses2010abundance}. Figure~\ref{fig:jupiter} shows the pressure-temperature profile, eddy diffusion coefficient, and the mixing ratios of \ce{CH4}, \ce{NH3}, and major photochemical gases of the test case. The atmospheric structure model adequately predicts the tropospheric temperature profile and the pressure level of the tropopause, but it cannot generate a temperature inversion in the middle atmosphere (Figure~\ref{fig:jupiter}, panel a). We have run the photochemical model with the pressure-temperature profile measured in Jupiter and the modeled pressure-temperature profile (i.e., without the temperature inversion) to see how much the photochemical gas mixing ratios change.
	
We find that the photochemical model can predict the mixing ratios of \ce{C2H6}, \ce{C2H2}, and \ce{C2H4} measured in Jupiter's stratosphere, and the modeled profile of \ce{HCN} is consistent with the upper limit in Jupiter's upper troposphere when the measured pressure-temperature profile is adopted (Figure~\ref{fig:jupiter}, panel c). The only exception is the \ce{C2H2} mixing ratio at $\sim1$ Pa, where the modeled mixing ratio is greater than the measured value by $2\sim3\sigma$. This less-than-perfect performance may be due to the lack of \ce{C3}, \ce{C4}, and higher hydrocarbons in our reaction network. For example, \cite{moses2005photochemistry} was able to fit the \ce{C2H2} mixing ratio at $\sim1$ Pa together with other mixing ratio constraints, with a more complete hydrocarbon reaction network and specific choices in the eddy diffusion coefficient profiles for Jupiter's stratosphere. In terms of nitrogen photochemistry, our photochemical model finds that \ce{NH3} is depleted by photodissociation to the cloud deck, and the vast majority of the net photochemical removal of \ce{NH3} becomes \ce{N2H4} and then condenses out. A small fraction becomes \ce{N2} and \ce{HCN}. The abundance of \ce{HCN} is low ($\sim10^{-9}$) in the troposphere due to the photolysis of \ce{NH3} and \ce{CH4} occurring at well separated pressure levels, and is limited by the cold trap near the tropopause (Figure~\ref{fig:jupiter}). These behaviors are qualitatively similar to the past models of Jupiter's nitrogen photochemistry \citep[][]{atreya1977distribution,kaye1983formation,kaye1983hcn,moses2010abundance}.

Figure~\ref{fig:jupiter} also indicates that adopting the modeled pressure-temperature profile that does not have a stratosphere, while preserving the overall behavior of the atmospheric photochemistry, would under-predict the mixing ratios of \ce{C2H6} and \ce{C2H2} by approximately half an order of magnitude. We use the atmospheric structure model in this study for speedy exploration of the main photochemical behavior, and one should keep this context in mind when interpreting the results shown in Section~\ref{sec:result}.

Another interesting point to make is that the quantum yield of \ce{H} in the photodissociation \ce{C2H2} has been convincingly measured to be 100\% by recent experiments \citep{lauter2002absolute}. When producing the models shown as the solid and dashed lines in Figure~\ref{fig:jupiter}, panel c, we have applied a quantum yield of 16\% so that the top-of-atmosphere rate of \ce{C2H2 + h$\nu$ -> C2H + H} would match with the models of \cite{gladstone1996hydrocarbon,moses2005photochemistry}. Revising the quantum yield to 100\%, as shown by the dotted lines in Figure~\ref{fig:jupiter}, panel c, slightly reduces the steady-state mixing ratio of \ce{C2H6} and reduces the mixing ratio of \ce{C2H2} and \ce{C2H4} by a factor of $\sim5$ in the lower stratosphere ($\sim10^3$ Pa). The photodissociation of \ce{C2H2} is the main source of \ce{H} in the lower stratosphere \cite[e.g.,][]{gladstone1996hydrocarbon} and thus its quantum yield is important for the hydrocarbon chemistry in the lower stratosphere. However, a quantum yield of 100\% would result in poor fits to the measured mixing ratios of \ce{C2H2} and \ce{C2H4}, and this potential discrepancy suggests that additional consideration of the atmospheric photochemistry of Jupiter might be warranted. We adopt the quantum yield of 100\% in the subsequent models.
	
\subsection{Planet Scenarios}
\label{sec:planet}
	
We use the temperate sub-Neptune K2-18~b as a representative case of temperate and gas-rich planets around M dwarf stars, and use the gas giants PH2~b and Kepler-167~e as the representative cases of temperate and cold planets around G and K stars (Table~\ref{tab:planet_par}). The results for K2-18~b are generally applicable to temperate (mini-)Neptunes of M dwarf stars such as the recently detected TOI-1231~b. For K2-18~b, the interior structures, thermochemical abundances, and atmospheric circulation patterns have been studied \citep{benneke2019water,madhusudhan2020interior,piette2020temperature,Blain2020,charney2020}, but the effects of atmospheric photochemistry remain to be studied. Kepler-167 e is considered a ``cold'' exoplanet because it only receives stellar irradiation $7.5\%$ of Earth's. The equilibrium cloud condensation model would predict NH$_3$ to condense in its atmosphere and form the uppermost cloud deck, below which NH$_4$SH solids form and scavenge sulfur from the above-cloud atmosphere. In the atmospheres of K2-18~b and PH2~b, only H$_2$O is expected to condense and forms the cloud deck -- and thus the physical distinction between ``temperate'' and ``cold''.

	\begin{deluxetable}{c|ccc}
		\tablecaption{Planetary parameters adopted in this study. \label{tab:planet_par}}
		\tablehead{
		\colhead{Planet} & \colhead{K2-18 b} & \colhead{PH2 b} & \colhead{Kepler-167 e} }
		\startdata
		$M_p$ (M$_{\oplus}$)  &   8.63\tablenotemark{1} & N/A & N/A \\
		$R_p$ (R$_{\oplus}$) &   2.61\tablenotemark{2} & 9.40\tablenotemark{3} & 9.96\tablenotemark{3} \\
		Insolation (Earth) & 1.0\tablenotemark{1} & 1.2\tablenotemark{4} & 0.075\tablenotemark{5}\\
		Stellar Type &  M & G & K  \\
		\enddata
		\tablecomments{$^1$\cite{benneke2019water}. $^2$\cite{cloutier2019confirmation}. $^3$\cite{berger2018revised}. $^4$\cite{wang2013planet}. $^5$\cite{kipping2016transiting}.}
	\end{deluxetable}

\begin{figure}
	\centering
	\includegraphics[width=0.45\textwidth]{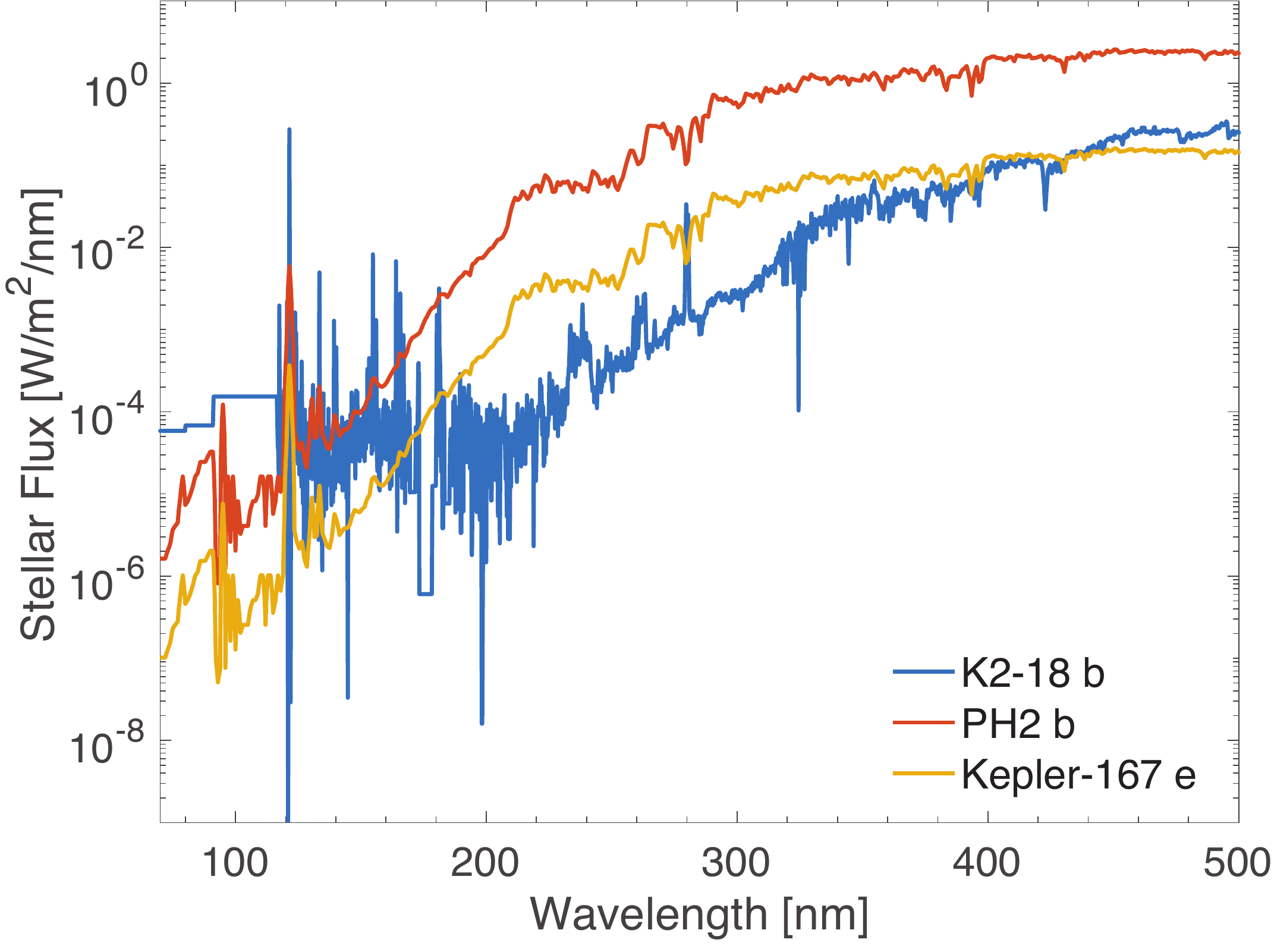}
	\caption{Incident stellar flux at the top of the atmospheres adopted in this study.}
	\label{fig:star}
	\end{figure}

The UV spectrum of K2-18 has not been measured and so we adopt that of GJ~176, a similar M dwarf star with the UV spectrum measured in the MUSCLES survey \citep{france2016muscles}. The reconstructed Ly-$\alpha$ flux of GJ~176 is similar to the measured flux of K2-18 \citep{dos2020high}. We adopt the UV spectrum of the Sun for the models of PH2~b and Kepler-167~e, even though Kepler-167 is a K star. Figure~\ref{fig:star} shows the incident stellar flux at the top of the atmospheres adopted in this study. K2-18~b, while having similar total irradiation as PH2~b, receives considerably less irradiation in the near-UV.
	
For these planets, we simulate H$_2$-dominated atmospheres having $1-100\times$ solar metallicities. The higher-than-solar metallicity scenario may be particularly interesting for sub-Neptunes like K2-18~b because of a proposed mass-metallicity relationship that posits a less massive planet should have a higher metallicity \citep{thorngren2016mass}. For PH2~b and Kepler-167 e, we assume as fiducial values a surface gravity of 25 m s$^{-2}$ and an internal heat flux that corresponds to $T_{\rm int}=100$ K, similar to the parameters of Jupiter. Changing the surface gravity to 100 m s$^{-2}$ results in slightly different cloud pressures and above-cloud abundance of gases on these planets, but do not change the qualitative behaviors of the atmospheric chemistry. For K2-18~b we assume an internal heat flux that corresponds to $T_{\rm int}=60$ K, similar to that of Neptune.

In the standard models, we assume that the dominant O, C, N, and S species are \ce{H2O}, \ce{CH4}, \ce{NH3}, and \ce{H2S} at the base of the photochemical domain. Gases and aerosols produced in the photochemical domain can be transported through the lower boundary, and thus the standard model setup implicitly assumes that thermochemical recycling in the deep troposphere effectively recycles the photochemical products into \ce{H2O}, \ce{CH4}, \ce{NH3}, and \ce{H2S}. Here we quantitatively assess how realistic this assumption is based on the quench-point theory \citep[e.g.,][]{visscher2011quenching,moses2013compositional,Hu2014B2014ApJ...784...63H,zahnle2014methane,tsai2018toward}. In that theory, the ``quench point'' is defined as the pressure level where the chemical lifetime of a gas equals the vertical mixing timescale (typically at the pressure of $10^7$ Pa or higher). The gas is close to thermochemical equilibrium at the quench point, and its mixing ratio is carried to the atmosphere above the quench point by vertical mixing.

\begin{figure}[!htbp]
\centering
\includegraphics[width=0.45\textwidth]{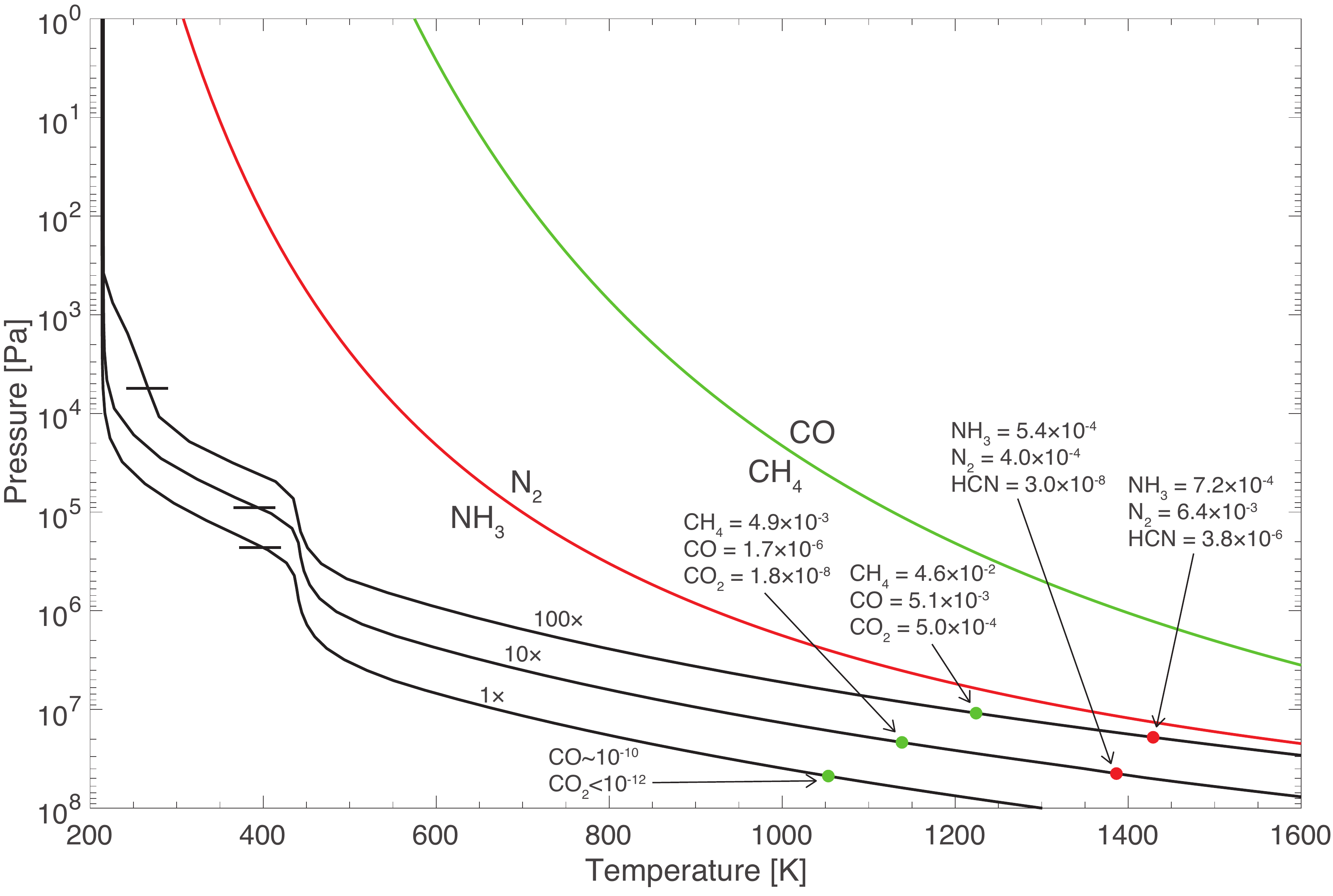}
\caption{
Pressure-temperature profiles of the temperate sub-Neptune K2-18~b for varied atmospheric metallicities and an internal heat flux of $T_{\rm int}=60$ K (similar to Neptune). The short horizontal bars show the lower boundary of the photochemical model (i.e., the pressure level 10-fold greater than the tropopause pressure). The green and red lines show the equal-abundance boundaries for major carbon and nitrogen gases in a solar-metallicity gas in thermochemical equilibrium, and the green and red dots show the expected quench point for CO and that for N$_2$ respectively. The equilibrium mixing ratios of major C and N molecules at the respective quench points are shown. 
}
\label{fig:quench_k2_18}
\end{figure}

\begin{figure}[!htbp]
\centering
\includegraphics[width=0.45\textwidth]{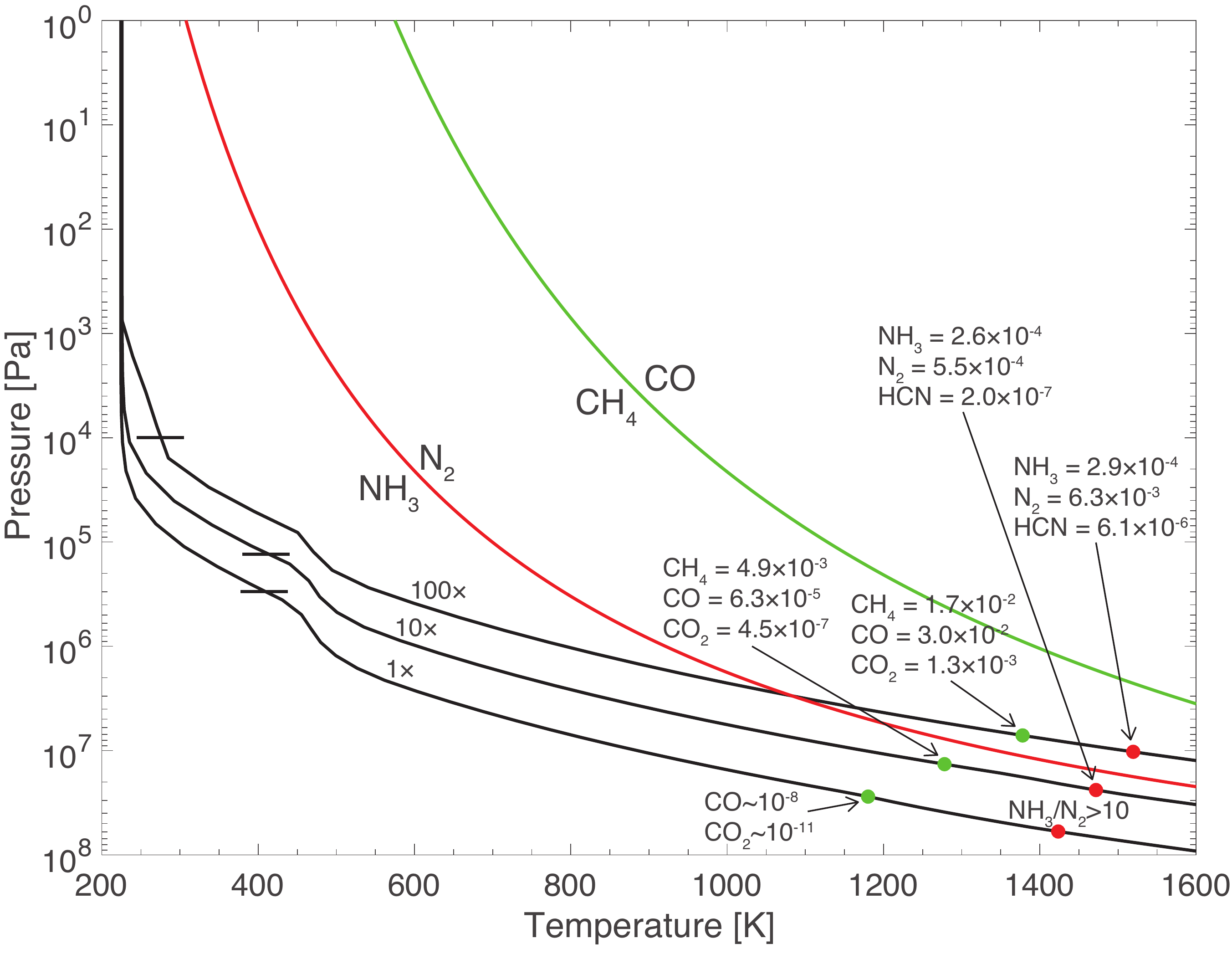}
\caption{
The same as Figure~\ref{fig:quench_k2_18} but for the planetary parameters of PH2~b and an internal heat flux of $T_{\rm int}=100$ K (similar to Jupiter).
}
\label{fig:quench_ph2b}
\end{figure}

\begin{figure}[!htbp]
\centering
\includegraphics[width=0.45\textwidth]{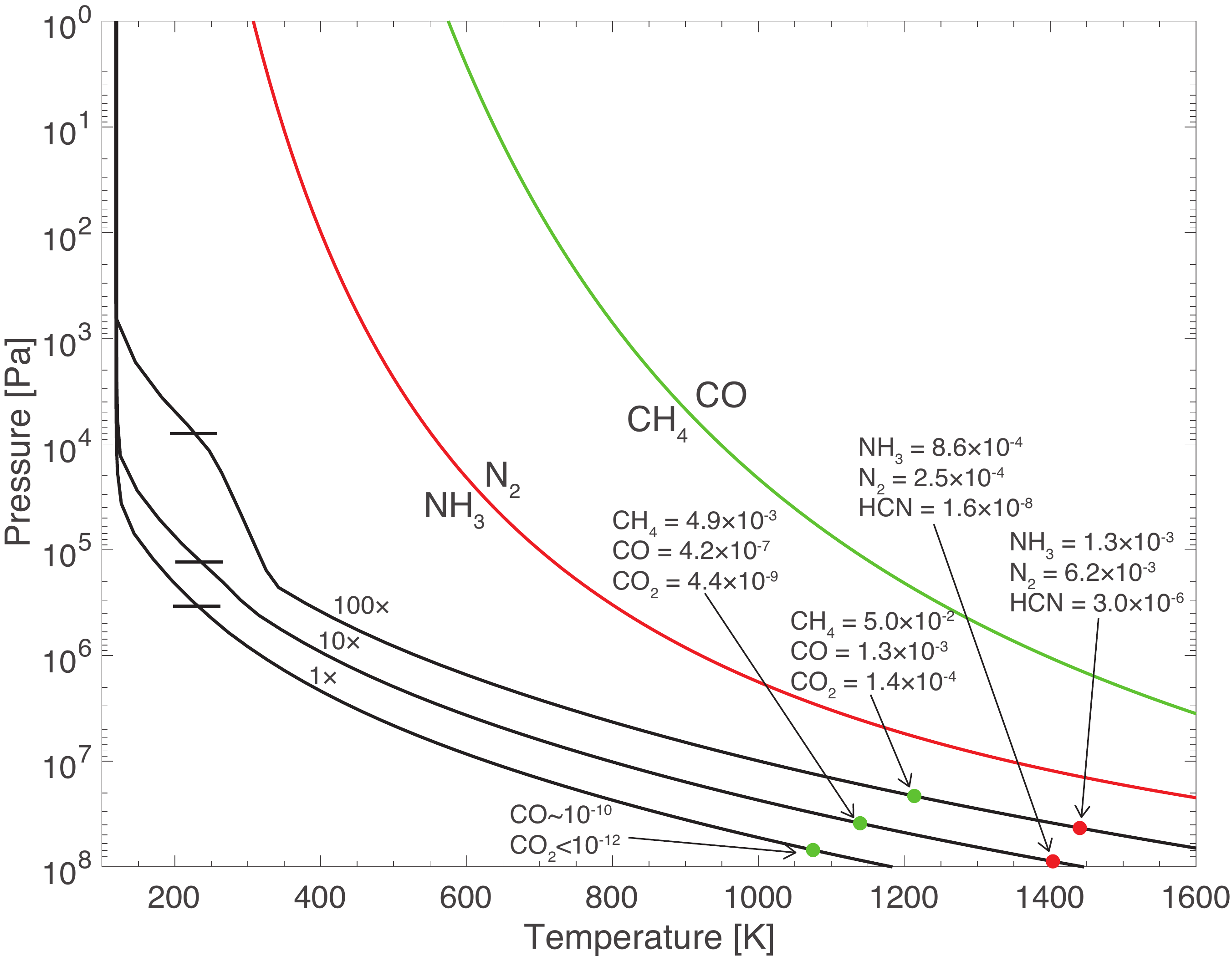}
\caption{
The same as Figure~\ref{fig:quench_k2_18} but for the planetary parameters of Kepler-167~e and an internal heat flux of $T_{\rm int}=100$ K (similar to Jupiter).
}
\label{fig:quench_kepler167e}
\end{figure}

Figures~\ref{fig:quench_k2_18} -- \ref{fig:quench_kepler167e} show the pressure-temperature profiles of the three planets calculated by the atmospheric structure model, and the mixing ratios of major \ce{C} and \ce{N} molecules at the respective quench points. We adopt the chemical lifetime of the \ce{CO <-> CH4} and \ce{N2 <-> NH3} conversions from \cite{zahnle2014methane} and estimate the eddy diffusion coefficient in the deep troposphere using the mixing-length theory in \cite{visscher2010deep}. The eddy diffusion coefficient depends on the assumed internal heat flux and has a typical value of $\sim10^4$ m$^2$ s$^{-1}$ at the pressure of $10^6-10^8$ Pa. The quench point of \ce{CO2} follows that of \ce{CO}, and similarly, that of \ce{HCN} occurs at a similar pressure and temperature as \ce{N2} \citep{zahnle2014methane,tsai2018toward}. The mixing ratios of gases at the quench points are calculated using the thermochemical equilibrium model of \cite{Hu2014B2014ApJ...784...63H}.

Figure~\ref{fig:quench_k2_18} -- \ref{fig:quench_kepler167e} show that a solar-metallicity atmosphere is likely deep in the \ce{CH4}- and \ce{NH3}-dominated regime at the quench points on all three planets. Specifically, we find the mixing ratio of \ce{CO} $\leq10^{-8}$, that of \ce{CO2} $\leq10^{-11}$, and the mixing ratio of \ce{NH3} greater than that of \ce{N2} by $>10$ folds. With  $10\times$solar metallicity, the atmosphere remains \ce{CH4}-dominated, but the mixing ratio of \ce{CO} transported from the deep troposphere can be on the order of $10^{-6}\sim10^{-5}$ and thus non-negligible. With the assumed internal heat flux and the modeled strength of deep tropospheric mixing, the mixing ratio of \ce{N2} can be comparable to that of \ce{NH3} at the quench point. As \ce{N2} does not have strong spectral features and is not a feedstock molecule for photochemistry, the effect of a hot interior would be mostly seen as a reduction of the mixing ratio of \ce{NH3}. The impact of the hot interior is the most significant in the $100\times$solar-metallicity atmosphere. Both \ce{CO} and \ce{CO2} have mixing ratios $>10^{-4}$ at the quench point, and in the hottest case (PH2~b), the mixing ratio \ce{CO} is greater than that of \ce{CH4}. For nitrogen, the mixing ratio of \ce{NH3} can be reduced by a factor of $10\sim100$ at the thermochemical equilibrium in the deep troposphere. 

As a general trend, a higher deep-atmosphere temperature favors \ce{CO}, \ce{CO2}, and \ce{N2}, and reduces the equilibrium abundance of \ce{NH3}. We have thus run variant models for the $10\times$ and $100\times$solar-metallicity cases, and used the mixing ratios of \ce{CH4}, \ce{CO}, \ce{CO2}, \ce{NH3}, and \ce{N2} at the quench points as shown in Figures~\ref{fig:quench_k2_18} -- \ref{fig:quench_kepler167e} as the lower-boundary conditions. Technically the mixing ratio of deep \ce{H2O} is also affected, but the photochemical models have lower boundaries that are well above the base of the water cloud, and are thus immune to small changes in the input water abundance. Also, we do not fix the lower-boundary mixing ratio of \ce{HCN} in these models, because the mixing ratio of \ce{HCN} at the quench point does not exceed the mixing ratio found by the photochemical models at the lower boundary in any case. We emphasize that specific quantities of the input gas abundance depend on the detailed thermal structure of the interior, which is related to the thermal history of the planet and exogenous factors like tidal heating, as well as the strength of vertical mixing in the interior \citep{fortney2020beyond}. For example, applying an internal heat flux that corresponds to $T_{\rm int}=30$ K (similar to Earth) largely restores the \ce{CH4} and \ce{NH3} dominance for the three planets. While these factors are likely uncertain for many planets to be observed, the standard and variant photochemical models presented in this paper give an account of the range of possible behaviors that manifest in the observable part of the atmosphere.
	
\section{Results} \label{sec:result}
	
\subsection{Main Behaviors of Atmospheric Chemistry} \label{sec:chem}

    \begin{figure*}
	\centering
	\includegraphics[width=0.65\textwidth]{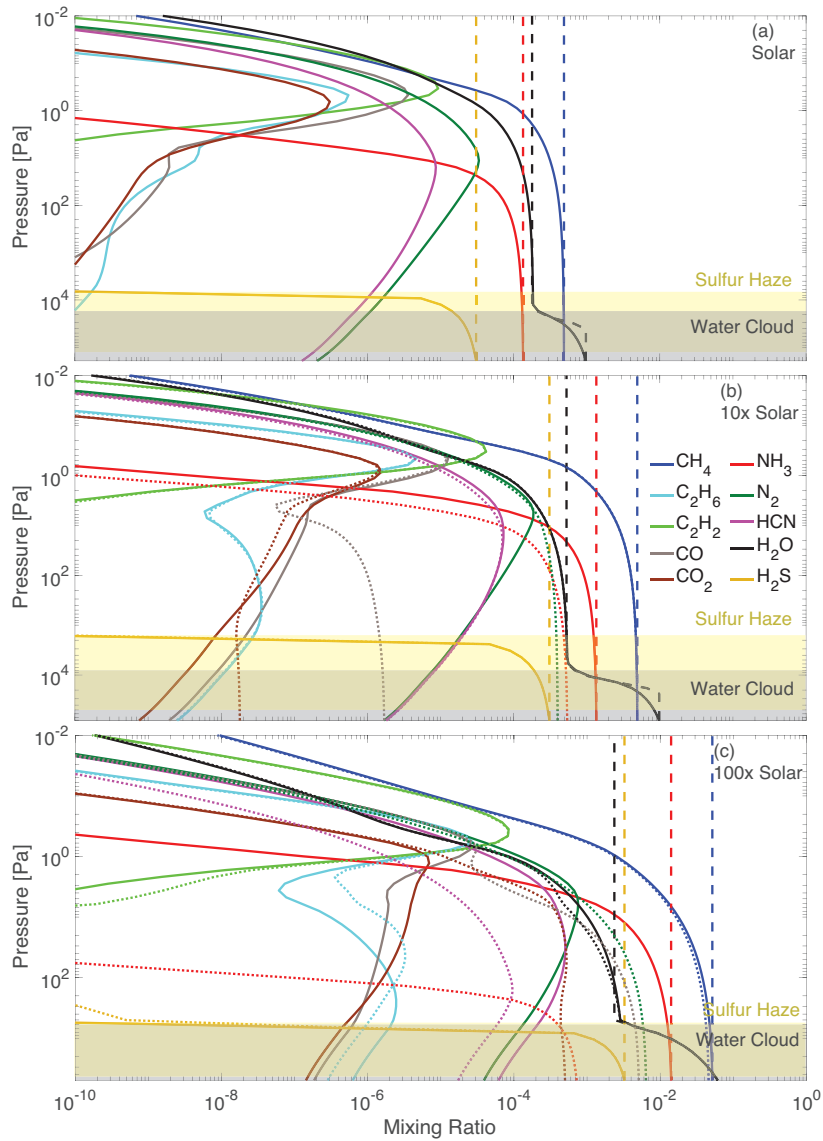}
	\caption{Modeled abundances of main gases and photochemical products in the temperate sub-Neptune K2-18~b for varied metallicities. Solid lines show the photochemical model results and the dashed lines show the equilibrium cloud condensation model results for comparison. Dotted lines in Panels (b) and (c) show the photochemical model results that adopt the quench-point abundances of \ce{CH4}, \ce{CO}, \ce{CO2}, \ce{NH3}, and \ce{N2} (Figure~\ref{fig:quench_k2_18}) at the lower boundary.}
	\label{fig:k218b_1}
	\end{figure*}
	
	\begin{figure*}
	\centering
	\includegraphics[width=0.65\textwidth]{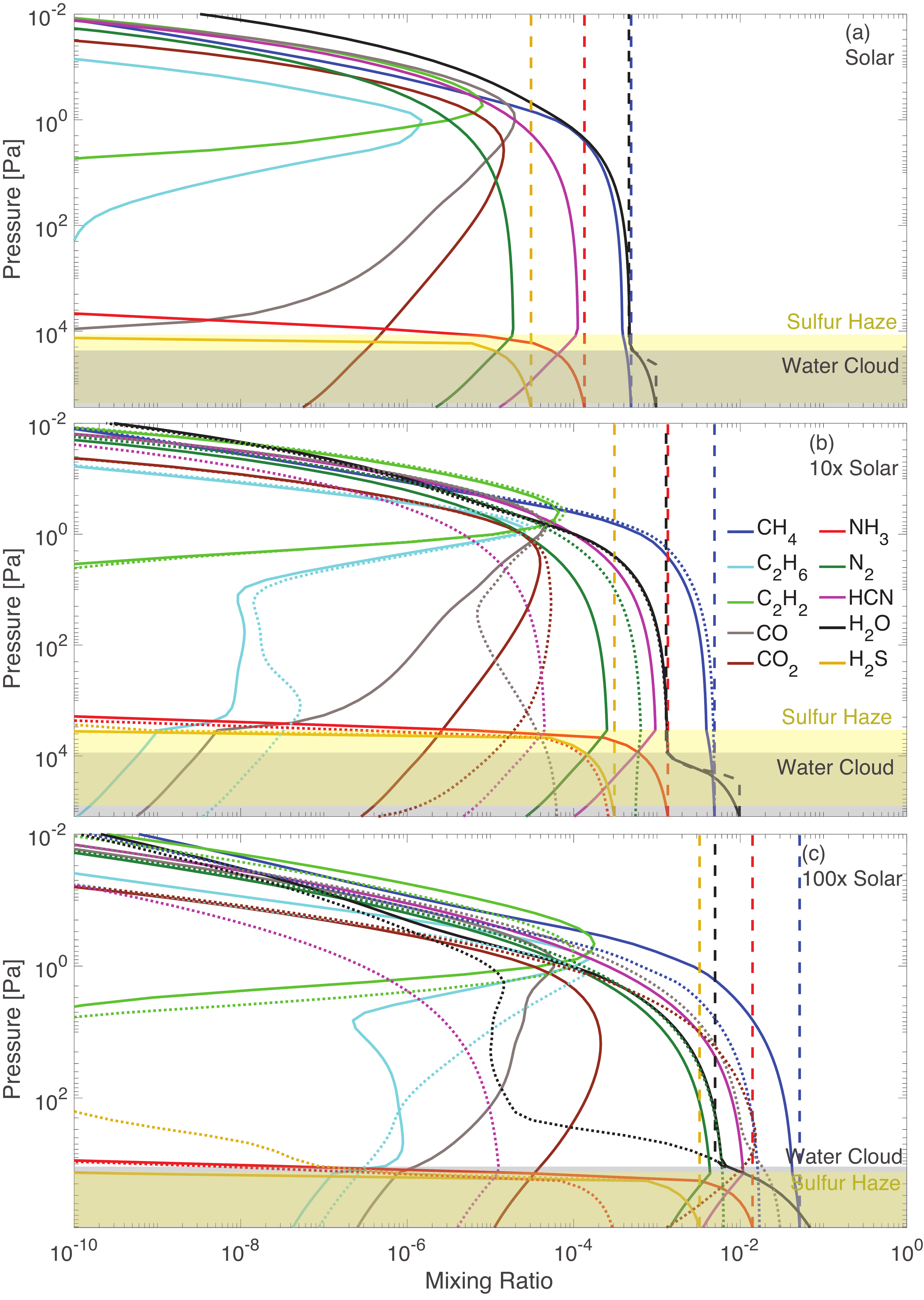}
	\caption{Modeled abundances of main gases and photochemical products in the temperate gas giant PH2~b for varied metallicities. Solid lines show the photochemical model results and the dashed lines show the equilibrium cloud condensation model results for comparison. Dotted lines in Panels (b) and (c) show the photochemical model results that adopt the quench-point abundances of \ce{CH4}, \ce{CO}, \ce{CO2}, \ce{NH3}, and \ce{N2} (Figure~\ref{fig:quench_ph2b}) at the lower boundary.}
	\label{fig:ph2b_1}
	\end{figure*}
	
    \begin{figure*}
	\centering
	\includegraphics[width=0.65\textwidth]{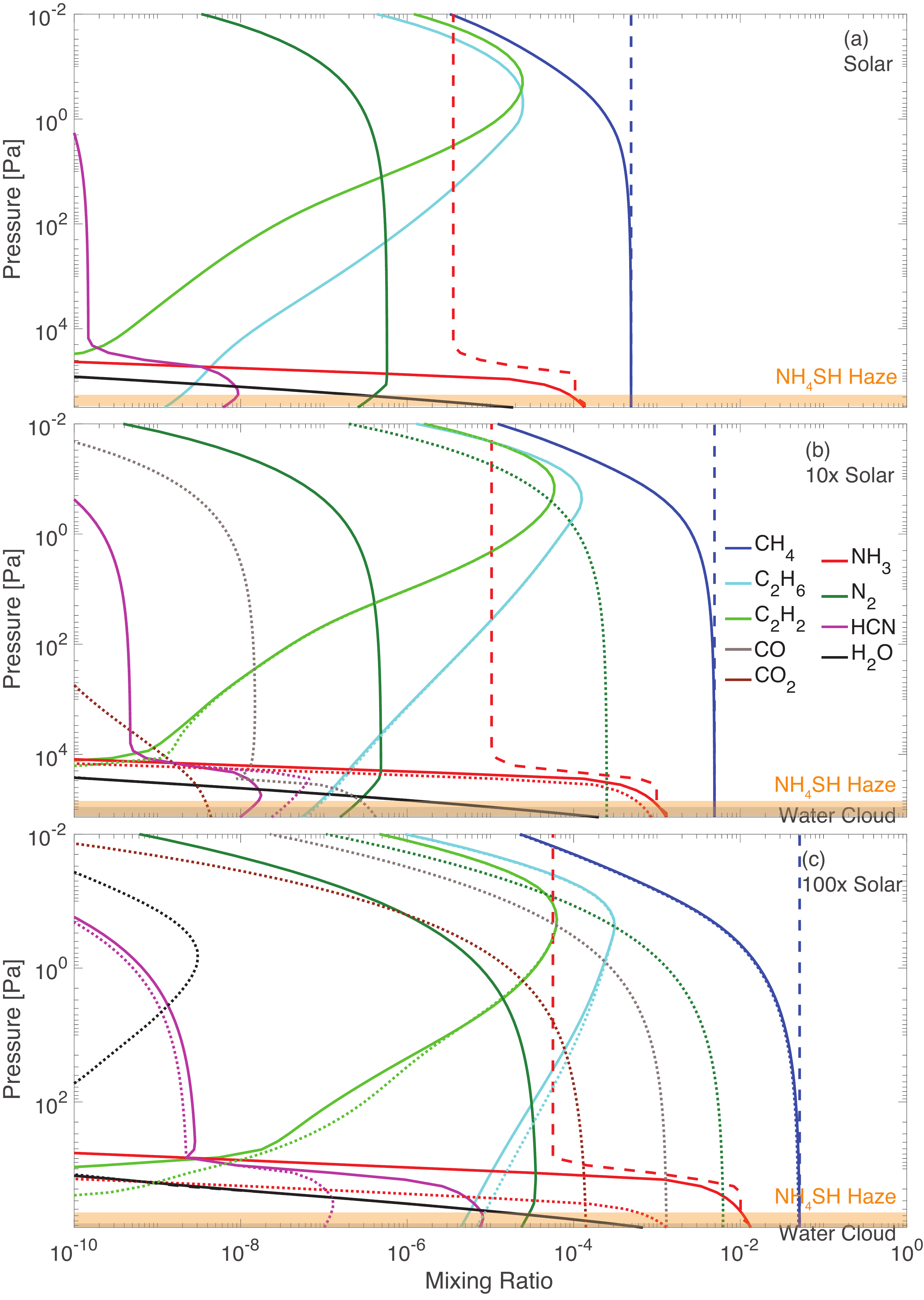}
	\caption{Modeled abundances of main gases and photochemical products in the cold gas giant Kepler-167~e for varied metallicities. Solid lines show the photochemical model results and the dashed lines show the equilibrium cloud condensation model results for comparison. Dotted lines in Panels (b) and (c) show the photochemical model results that adopt the quench-point abundances of \ce{CH4}, \ce{CO}, \ce{CO2}, \ce{NH3}, and \ce{N2} (Figure~\ref{fig:quench_kepler167e}) at the lower boundary.}
	\label{fig:kepler167e_1}
	\end{figure*}
	
The abundance profiles of the equilibrium gases (\ce{H2O}, \ce{CH4}, \ce{NH3}, and \ce{H2S}) and their main photochemical products are shown in Figures~\ref{fig:k218b_1} -- \ref{fig:kepler167e_1}. In all models, we find that the main photochemical products are \ce{C2H6}, \ce{C2H2}, \ce{CO}, \ce{CO2}, \ce{N2}, \ce{HCN}, \ce{N2O}, and elemental sulfur haze, but the abundance of these species and their pressure dependency differ significantly from model to model.
	
\subsubsection{K2-18~b: a temperate planet around an M star}
	
For K2-18~b, our model predicts that water condenses to form a cloud at the pressure of $\geq10^4$ Pa for the solar and $10\times$solar cases, and at the pressure of $\sim10^3$ Pa for the $100\times$solar case. Above the cloud, the mixing ratio of water is depleted by approximately one order of magnitude, but not totally depleted. The pressure of cloud for the $100\times$solar abundance case we model is consistent with predictions of a non-gray radiative-equilibrium model and a 3D climate model, but those models do not predict a water cloud for the solar and $10\times$solar abundance \citep{Blain2020,charney2020}. Given the small degree of water depletion found in our models, this discrepancy does not lead to substantial errors in the results of the above-cloud photochemistry.
	
Both \ce{CH4} and \ce{H2O} are photodissociated at the pressure of approximately 0.1 -- 1 Pa. The photodissociation results in the formation of \ce{C2H6}, \ce{C2H2}, \ce{CO}, and \ce{CO2}. \ce{C2H2} has a high mixing ratio at the pressure where the photodissociation takes place but is quickly depleted towards higher pressures. In the middle atmosphere ($\sim10$ -- $10^3$ Pa), \ce{CO}, \ce{CO2}, and \ce{C2H6} can have a mixing ratio of $\sim1$ parts-per-million (ppm) for the $100\times$solar abundance case, and the mixing ratio of these photochemical gases is $<1$ ppm for lower metallicities. When the deep tropospheric source of \ce{CO} and \ce{CO2} is applied to the bottom of the photochemical domain, the mixing ratio of \ce{CO} at $10^2$ Pa is $\sim1$ ppm for the $10\times$solar cases, but it can reach $\sim4000$ ppm for the $100\times$solar case. The mixing ratio of \ce{CO2} at $10^2$ Pa can reach $\sim500$ ppm for the $100\times$solar case.
	
\ce{NH3} is photodissociated at the pressure of $1\sim10$ Pa. The photodissociation results in the formation of \ce{N2} and \ce{HCN} with similar yields. The mixing ratio of \ce{HCN} at $\sim10^2$ Pa is $\sim6$, 50, and 500 ppm for the solar, $10\times$solar, and $100\times$solar abundance cases, respectively. If the mixing ratio of \ce{NH3} in the deep troposphere is applied to the bottom of the photochemical domain, the resulting mixing ratio of \ce{HCN} does not change significantly in the $10\times$solar case but decreases to $\sim100$ ppm in the $100\times$solar case.
	
Lastly, \ce{H2S} is photodissociated at approximately the same pressure as the water cloud. The photodissociation leads to the formation of elemental sulfur (\ce{S8}) haze, as predicted previously \citep{zahnle2016photolytic}. The haze layer extends to an altitude only slightly higher than the water cloud deck.
	
\subsubsection{PH2~b: a temperate planet around a G/K star}
	
PH2~b has a slightly higher insolation and temperature than K2-18~b, but it receives much more near-UV irradiation (Figure~\ref{fig:star}). The water condensation and small degree of depletion above the cloud, as well as the photodissociation of \ce{H2S} and the location of the sulfur haze layer, are similar to those predicted for K2-18~b.
	
\ce{CH4} is photodissociated at the pressure of 0.1 -- 1 Pa, and \ce{H2O} is photodissociated at 1 -- 10 Pa. The main products of these photodissociations are still \ce{C2H6}, \ce{C2H2}, \ce{CO}, and \ce{CO2}. Instead of \ce{CO} in the case of K2-18~b, \ce{CO2} is the most abundant photochemical gas in the middle atmosphere ($\sim10$ -- $10^3$ Pa), and its mixing ratio is 2 -- 10 ppm, 5 -- 40 ppm, and 40 -- 200 ppm for the solar, $10\times$solar, and $100\times$solar abundance cases, respectively. The mixing ratio of \ce{CO} is less by approximately one order of magnitude, and that of \ce{C2H6} is $\sim1$ ppm for the $100\times$solar case and $<1$ ppm for lower metallicities.
	
As a striking difference from the M star case (K2-18~b), \ce{NH3} is fully depleted by photodissociation above the water cloud deck. The mixing ratio of \ce{NH3} in the middle atmosphere is minimal. The photodissociation also leads to the formation of \ce{N2} and \ce{HCN}, with \ce{HCN} being the most abundant photochemical product. The mixing ratio of \ce{HCN} in the middle atmosphere reaches $\sim100$, 700, and 10,000 ppm for the solar, $10\times$solar, and $100\times$solar abundance cases, respectively.

With a Jupiter-like internal heat flux, the equilibrium chemistry in the deep troposphere may substantially change the chemical composition in the photochemical domain. In the $10\times$solar cases, the mixing ratio of \ce{CO} in the middle atmosphere can reach $\sim10$ ppm and that of \ce{CO2} $\sim60$ ppm. \ce{HCN} would no longer be the most abundant nitrogen product, and its mixing ratio in the middle atmosphere can be reduced to $\sim40$ ppm. In the $100\times$solar cases, both \ce{CO} and \ce{CO2} can have very high mixing ratios ($>10^{-2}$, and on the same order of \ce{CH4}) in the middle atmosphere, and the above-cloud \ce{H2O} would be consumed by photochemistry and have a mixing ratio of $\sim10$ ppm at $10^2$ Pa. The mixing ratio of \ce{HCN} would be further reduced to $\sim10$ ppm, while still marginally greater than the mixing ratio at the quench point.
	
\subsubsection{Kepler-167~e: a cold planet around a G/K star}
	
The atmosphere of Kepler-167~e is much colder than that of K2-18~b or PH2~b, and its atmospheric chemistry is more akin to that of Jupiter \citep{gladstone1996hydrocarbon,moses2005photochemistry,atreya1977distribution,kaye1983formation,kaye1983hcn,moses2010abundance}. Both \ce{H2O} and \ce{H2S} are fully depleted by condensation or \ce{NH4SH} formation, and the uppermost cloud predicted by the atmospheric structure model is \ce{NH3} ice. However, the steady-state results of the photochemical model indicate that photodissociation of \ce{NH3} should deplete the \ce{NH3} ice cloud. \ce{NH3} is photochemically depleted to the pressure of $7\times10^4$ -- $10^3$ Pa from the solar to $100\times$solar abundance cases. The main product of the photodissociation that can accumulate in the middle atmosphere is \ce{N2}, while the mixing ratios of \ce{HCN} and \ce{N2H4} are limited by condensation. The mixing ratio of \ce{HCN} can reach $>1$ ppm below the condensation level in the $100\times$solar case.

The main photochemical gases of carbon are \ce{C2H6} and \ce{C2H2}, with no \ce{CO} or \ce{CO2} at appreciable mixing ratios. While the mixing ratio of \ce{C2H2} strongly peaks at $0.1$ Pa, where the photodissociation of \ce{CH4} takes place, the mixing ratio of \ce{C2H6} can be significant in the middle atmosphere. At $10^2$ Pa, the mixing ratio of \ce{C2H6} is $\sim2$, 4, and 30 ppm for the solar, $10\times$solar, and $100\times$solar abundance cases, respectively. If the deep tropospheric source of \ce{CO} and \ce{CO2} is applied to the bottom of the photochemical domain, they can have substantial mixing ratios in the $100\times$solar case, while the mixing ratio of \ce{C2H6} is not strongly impacted.
	
\subsection{Photochemical Depletion of NH$_3$} \label{sec:pres}
	
From Figures \ref{fig:k218b_1} -- \ref{fig:kepler167e_1}, we see that \ce{NH3} is depleted to the cloud deck in temperate and cold planets around G/K stars but remain intact in the middle atmosphere of temperate and cold planets around M stars. This finding is significant because it implies that \ce{NH3} should be detectable on temperate planets around M stars but not around G/K stars (see Section~\ref{sec:spec}).
    
The root cause of this different behavior is the M stars (represented by GJ~176 here) emit substantially lower irradiation at the near-UV wavelengths than the G/K stars (represented by the Sun here, Figure~\ref{fig:star}). The radiation that dissociates \ce{NH3} in the H$_2$-dominated atmosphere is the radiation that is not absorbed by the typically more abundant \ce{CH4} and \ce{H2O}. \ce{NH3} has a dissociation limit at $\sim230$ nm while \ce{CH4} at $\sim150$ nm and \ce{H2O} at $\sim240$ nm, but the cross section and the shielding effect of \ce{H2O} is small $>200$ nm \citep{hu2012photochemistry,ranjan2020photochemistry}. \ce{C2H2} also absorbs photons up to $\sim230$ nm but it typically does not strongly interfere with the \ce{NH3} photodissociation due to its relatively low abundance. Thus, photons in 200 -- 230 nm are the most relevant for the photodissociation of \ce{NH3} in K2-18~b and PH2~b, and photons in 150 -- 230 nm are the most relevant for Kepler-167 e. Having similar bolometric irradiation, the photon flux in 200 -- 230 nm received by PH2~b is more than that received by K2-18~b by $>2$ orders of magnitude (Figure~\ref{fig:star}). The photon flux received by Kepler-167~e is one-order-of-magnitude more than K2-18~b, and the removal of \ce{NH3} by condensation further pushes down the pressure of photochemical depletion (see below).
    
\subsubsection{Criterion of Photochemical Depletion}
\label{sec:criterion}
    
How does the photon flux control the pressure of photochemical depletion? Guided by the numerical results, here we develop a simple theory that estimates the pressure of photochemical depletion. Assuming that photodissociation is the only process that removes \ce{NH3} with no recycling or production, its mixing ratio profile at the steady state should obey the following differential equation:
    \begin{equation}
        \frac{d}{d z}\bigg{(}KN\frac{d f}{d z}\bigg{)} = fNJ,\label{eq:general}
    \end{equation}
    where $z$ is altitude, $K$ is the eddy diffusion coefficient, $N$ is the total number density of the atmosphere, $f$ is the mixing ratio, and $J$ is the photodissociation rate (often referred to as the ``J-value'' in the atmospheric chemistry literature). The number density has a scale height of $H$, and the equation can be rewritten as
    \begin{equation}
        \frac{d^2f}{dz^2}-\frac{1}{H}\frac{df}{dz}-\frac{J}{K}f=0. \label{eq:nh3_0}
    \end{equation}
    Assuming $J$, $H$, and $K$ to be a constant with respect to $z$, the equation above has the analytical solution as
    \begin{equation}
        f = f_0 \exp\bigg{(}\frac{z}{2}\bigg{(}\frac{1}{H}-\sqrt{\frac{1}{H^2}+\frac{4J}{K}}\bigg{)}\bigg{)}\equiv f_0\exp\bigg{(}-\frac{\alpha z}{H}\bigg{)}, \label{eq:nh3_1}
    \end{equation}
    where $f_0$ is the mixing ratio at the pressure of photochemical depletion ($z=0$ for simplicity), and $\alpha$ is
    \begin{equation}
        \alpha = \frac{1}{2}\bigg{(}\sqrt{1+\frac{4JH^2}{K}}-1\bigg{)}.\label{eq:nh3_2}
    \end{equation}
    Therefore, when the product $4JH^2/K$ is small, $\alpha\rightarrow0$ and the mixing ratio profile is close to a constant; and when $4JH^2/K$ is large, $\alpha$ can be $\gg1$ and thus the mixing ratio drops off very quickly. This explains the vertical profiles of \ce{NH3} seen in Figures \ref{fig:k218b_1}-\ref{fig:kepler167e_1}.
    
    Going back to Equation (\ref{eq:general}), which can be integrated from the pressure of photochemical depletion to the top of the atmosphere, as
    \begin{equation}
    KN\frac{df}{dz}\arrowvert_{z=\infty} - KN\frac{df}{dz}\arrowvert_{z=0} = \int_{0}^{\infty}n(z)J(z)dz,\label{eq:int}
    \end{equation}
    where $n\equiv fN$ is the number density of \ce{NH3}. Assuming that the photoabsorption of \ce{NH3} itself is the sole source of opacity, $J$ can be expressed as
    \begin{equation}
        J(z) = J_{\infty} \exp(-\sigma\int_{z}^{\infty}n(z')dz'), \label{eq:j}
    \end{equation}
    where $J_{\infty}$ is the top-of-atmosphere J-value and $\sigma$ is the mean cross section of \ce{NH3}. The differential of Equation (\ref{eq:j}) is 
    \begin{equation}
        \frac{dJ}{dz}=\sigma nJ. \label{eq:dj}
    \end{equation}
    Combining Equations (\ref{eq:int}) and (\ref{eq:dj}), and recognizing $df/dz$ vanishes at $z=\infty$, we obtain
    \begin{equation}
        - KN\frac{df}{dz}\arrowvert_{z=0} = \frac{1}{\sigma}\int_0^{\infty}\frac{dJ}{dz}dz = \frac{J_{\infty}-J(z=0)}{\sigma}.
    \end{equation}
    With $J(z=0)\sim0$ (i.e., the J-value immediately below the pressure of photochemical depletion is minimal), and $J_{\infty}=\sigma I$, where $I$ is the photon flux at the top of the atmosphere, we obtain
    \begin{equation}
        -KN\frac{df}{dz}\arrowvert_{z=0}=I.\label{eq:flux}
    \end{equation}
    Note that to derive Equation (\ref{eq:flux}), no specific profiles for $J$ or $n$ ($f$) need to be assumed.
    
    The physical meaning of Equation (\ref{eq:flux}) is that the number of \ce{NH3} molecules that diffuse through the pressure of photochemical depletion should be equal to the number of photons received at the top of the atmosphere. This physical condition would become evident if one regards the column of \ce{NH3} above the pressure of photochemical depletion as a whole and recognizes that one photon dissociates one molecule. To the extent that the photoabsorption of \ce{NH3} itself is the dominant source of opacity, the criterion expressed by Equation (\ref{eq:flux}) does not depend on the mean cross section. Similarly, the criterion will be applicable to any molecule subject to photodissociation in a wavelength range largely free of interference by other molecules.
    
    It should be noted that Equation~(\ref{eq:flux}) cannot be derived by requiring the pressure of photochemical depletion to occur roughly at the optical depth of unity for the photodissociating radiation. This is because the mixing ratio profile in Equation~(\ref{eq:nh3_1}) is valid only locally and depends on $J$, which in turns depends on the vertical profile of the mixing ratio. As such, one cannot integrate Equation~(\ref{eq:nh3_1}) directly to find the pressure of photochemical depletion, and the optical-depth-of-unity condition is not as predictive as Equation~(\ref{eq:flux}).
    
    The left-hand side of Equation (\ref{eq:flux}) can be evaluated locally using Equation (\ref{eq:nh3_1}), and Equation (\ref{eq:flux}) becomes
    \begin{equation}
        \frac{\alpha KN_0f_0}{H} = I, \label{eq:criterion}
    \end{equation}
    where $N_0$ and $f_0$ is the total number density and the mixing ratio at the pressure of photochemical depletion. The pressure is thus $P_0=N_0k_bT$ where $k_b$ is the Boltzmann constant and $T$ is temperature. $\alpha$ can be evaluated with Equation (\ref{eq:nh3_2}) for a J value that corresponds to 5\% of the top-of-atmosphere value. Equation (\ref{eq:criterion}) thus provides a closed-form criterion that determines the pressure of photochemical depletion, and explains why the pressure of photochemical depletion is sensitive to the top-of-atmosphere flux of photons that drive photodissociation.
        
    \begin{figure}
	\centering
	\includegraphics[width=0.45\textwidth]{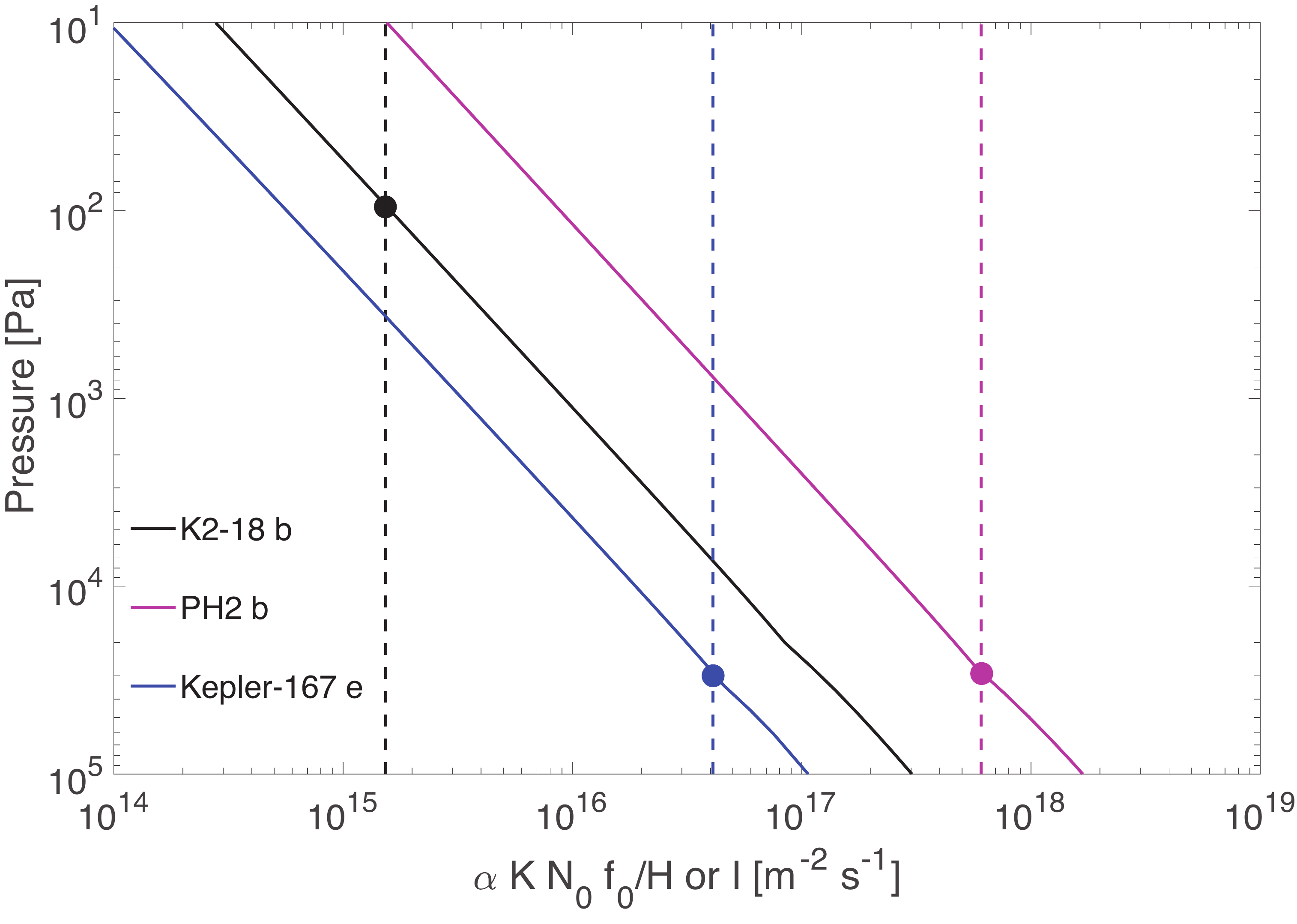}
	\caption{Pressure of photochemical depletion of \ce{NH3} predicted by the criterion in Equations~(\ref{eq:flux}) and (\ref{eq:criterion}). We compare the left-hand side (solid line) and right-hand side (dashed line) of Equation~(\ref{eq:criterion}), assuming a solar-abundance atmosphere. Where the solid line and the dashed line meet defines the pressure of photochemical depletion.}
	\label{fig:depth}
	\end{figure}
	
	Figure~\ref{fig:depth} shows both sides of Equation (\ref{eq:criterion}) for the three planets modeled assuming a solar-abundance atmosphere. We can see that the pressure of photochemical depletion implied by Equation (\ref{eq:criterion}) for K2-18~b is $\sim100$ Pa, consistent with the pressure where the photochemical model starts to substantially deviate from the equilibrium cloud condensation model (Figure~\ref{fig:k218b_1}). Figure \ref{fig:k218b_1} also shows that the mixing ratio \ce{NH3} decreases very slowly near the pressure of the photochemical depletion, but the decrease becomes faster for lower pressures, where the J value and the $4JH^2/K$ product become greater (see Equations~\ref{eq:nh3_1} and \ref{eq:nh3_2}). The mixing ratio of \ce{NH3} eventually drops below $10^{-6}$ at the pressure lower than the pressure of photochemical depletion by approximately one order of magnitude. The pressures of photochemical depletion implied by Equation (\ref{eq:criterion}) for PH2~b and Kepler-167~e are close to or below the cloud deck (i.e., $10^4-10^5$ Pa), which is consistent with numerical finding that \ce{NH3} is photodissociated to the cloud deck on these planets (Figures~\ref{fig:ph2b_1} and \ref{fig:kepler167e_1}). Therefore, although Equation~(\ref{eq:criterion}) cannot replace the full photochemical calculation due to the underlying assumptions (e.g., no recycling or production, self-shielding only), it provides a guiding estimate of whether a gas is likely depleted by photodissociation in the middle atmosphere.
    
\subsubsection{Sensitivity to the eddy diffusion coefficient}

    \begin{figure*}
	\centering
	\includegraphics[width=0.6\textwidth]{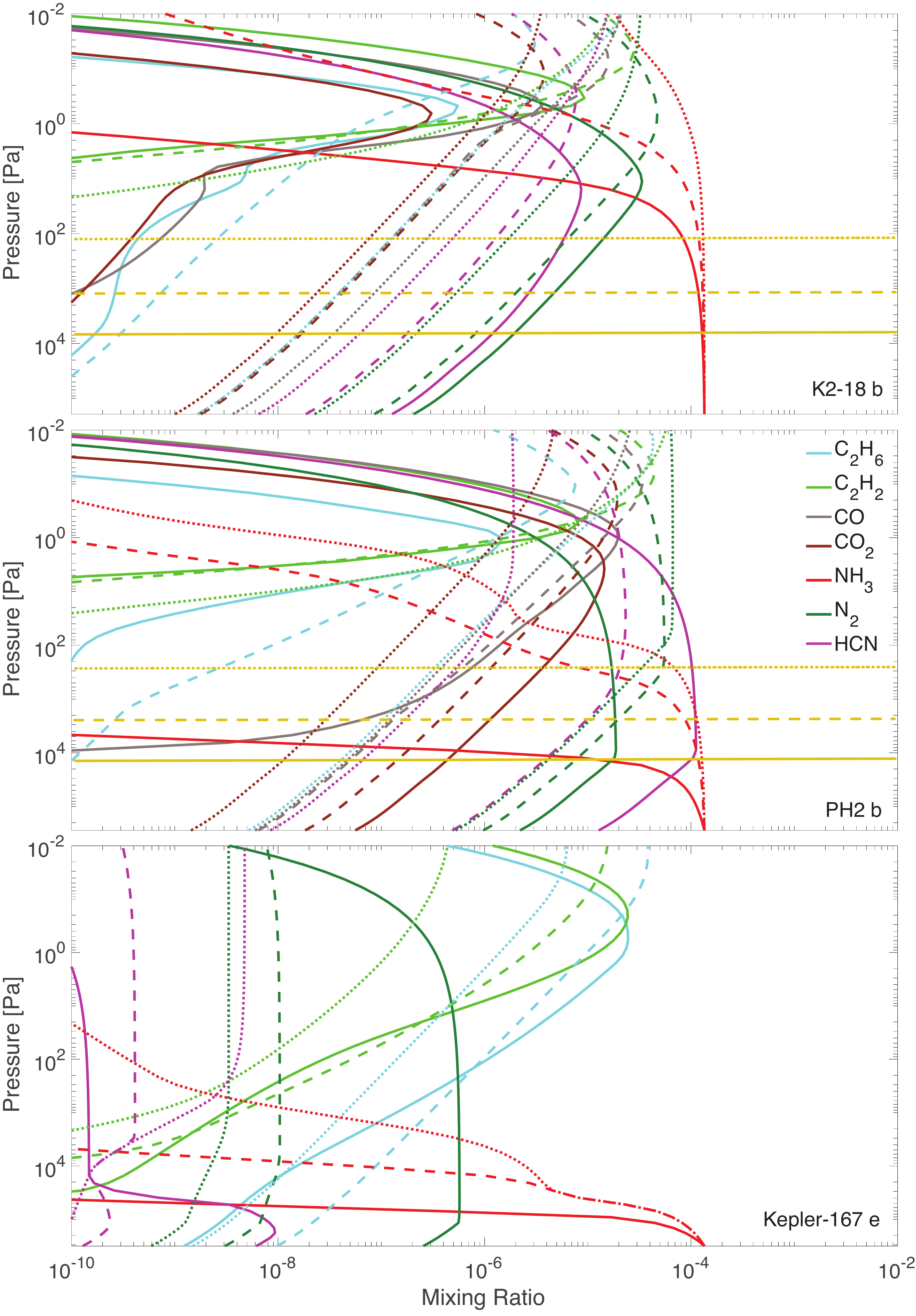}
	\caption{Sensitivity of the abundance profiles of \ce{NH3} and main photochemical gases to the eddy diffusion coefficient. The profiles of \ce{H2O} and \ce{CH4} are not shown because their abundance in the middle atmosphere is not sensitive to the eddy diffusion coefficient. The horizontal orange lines show the top of the sulfur haze layer. The solid lines show the standard model, and the dashed and dash-dot lines show the models with 10-fold and 100-fold greater eddy diffusion coefficients, respectively. These models assume the solar abundance. A greater eddy diffusion coefficient causes the photodissociation of \ce{NH3} to occur at a lower pressure.}
	\label{fig:k}
	\end{figure*}
	
From the criterion of photochemical depletion (Equation~\ref{eq:criterion}), we see that when the eddy diffusion coefficient increases, the pressure of photochemical depletion decreases. In other words, stronger mixing would sustain a photodissociated gas (e.g., \ce{NH3}) to a lower pressure or higher altitude. We have used the photochemical model to conduct a sensitivity study of the eddy diffusion coefficient, and the results confirm this understanding (Figure~\ref{fig:k}). The most significant sensitivity happens with PH2~b: the standard model predicts the photodissociation would deplete \ce{NH3} to the cloud deck, while with a 10-fold or 100-fold greater eddy diffusion coefficient, \ce{NH3} would be depleted at $10^2\sim10^3$ Pa. For Kepler-167~e, with a 10-fold or 100-fold greater eddy diffusion coefficient, photodissociation can no longer deplete the \ce{NH3} ice cloud, while the mixing ratio of \ce{NH3} in the middle atmosphere remains small due to condensation and photodissociation above the cloud deck.
	
The top of the sulfur haze layer moves up in the atmosphere when the eddy diffusion coefficient increases. For both K2-18~b and PH2~b, the top of the sulfur haze would be at $\sim10^3$ Pa and $10^2$ Pa for 10-fold and 100-fold greater eddy diffusion coefficient (Figure \ref{fig:k}). A haze layer that extends to $10^2$ Pa would greatly interfere with transmission spectroscopy and also affect the spectra of the reflected starlight (see Section \ref{sec:spec}). This trend is consistent with the findings of \cite{zahnle2016photolytic} and is produced by two effects acting together. First, the pressure of photochemical depletion of \ce{H2S}, the feedstock of sulfur haze, decreases for a greater eddy diffusion coefficient. Second, a stronger eddy diffusion helps increase the lifetime of haze particles against falling \citep[see the formulation in][]{hu2012photochemistry}. For PH2~b, the extended sulfur haze layer further keeps \ce{NH3} from photochemical depletion by absorbing the ultraviolet photons that can dissociate \ce{NH3}.
	
The sensitivity of main photochemical gases' abundance to the eddy diffusion coefficient is complex (Figure \ref{fig:k}), which indicates several factors at work. For \ce{N2} and \ce{HCN} (the dominant photochemical gases of nitrogen), their mixing ratios at the lower boundary decrease with the eddy diffusion coefficient. This is because, in our model, gases move across the lower boundary at a velocity that is proportional to the eddy diffusion coefficient, and the loss to the lower boundary is the main loss mechanism for both \ce{N2} and \ce{HCN}. Their mixing ratios in the middle atmosphere do not necessarily follow the same trend as that also depends on the photochemical production (see Section~\ref{sec:nh3}). The abundance of the photochemical gases of carbon does not depend on the eddy diffusion coefficient monotonically, and this is because the formation rates of \ce{CO}, \ce{CO2}, and \ce{C2H6} largely depend on the abundance of \ce{H}, \ce{OH}, and \ce{O}, which is in turn controlled by the full chemical network involving the photodissociation of \ce{CH4}, \ce{H2O}, and \ce{NH3} (see Section~\ref{sec:ch4}). For example, in K2-18~b with the solar metallicity, both 
\ce{CO} and \ce{CO2} have very small mixing ratios in the middle atmosphere in the standard case; the two would be substantially more abundant in the middle atmosphere with a 10-fold greater eddy diffusion coefficient, and \ce{CO} would become more abundant than \ce{CO2} with a 100-fold greater eddy diffusion coefficient. These examples highlight the richness and complexity of atmospheric photochemistry in temperate and cold planets.
	 
\subsection{Photolysis of NH$_3$ in the Presence of CH$_4$} \label{sec:nh3}
	
A common phenomenon that emerges from the photochemical models is the synthesis of \ce{HCN} in temperate and \ce{H2}-rich exoplanets. The photodissociation of \ce{NH3} in Jupiter leads to \ce{N2} but not significant amounts of \ce{HCN}, and this is mainly because \ce{NH3} is dissociated at much higher pressures than \ce{CH4} \citep[e.g.,][]{atreya1977distribution,kaye1983formation,kaye1983hcn,moses2010abundance}. \ce{HCN} in Titan's \ce{N2}-dominated atmosphere mainly comes from the reactions between atomic nitrogen and hydrocarbons and the associated chemical network \citep{yung1984photochemistry, lavvas2008coupling2, krasnopolsky2014chemical, vuitton2019simulating}. Similar processes, as well as the reactions between \ce{CH} and \ce{NO}/\ce{N2O} may also lead to formation of \ce{HCN} on early Earth or rocky exoplanets with \ce{N2}-dominated atmospheres irradiated by active stars \citep{airapetian2016prebiotic,rimmer2019hydrogen}. In addition, the formation of \ce{HCN} has been commonly found in warm and hot \ce{H2}-rich exoplanets \citep[e.g.,][]{moses2011disequilibrium,line2011thermochemical,venot2012chemical,agundez2014puzzling,molliere2015model,moses2016composition,blumenthal2018comparison,kawashima2018theoretical,molaverdikhani2019cold,hobbs2019chemical,lavvas2019photochemical}, and the mechanisms identified include quench kinetics \citep{moses2011disequilibrium,venot2012chemical,agundez2014puzzling} and photochemistry \citep{line2011thermochemical,kawashima2018theoretical,hobbs2019chemical}. Here we show that \ce{HCN} can also build up to significant amounts in temperate exoplanets with \ce{H2}-dominated atmospheres.
	
    \begin{figure}
	\centering
	\includegraphics[width=0.45\textwidth]{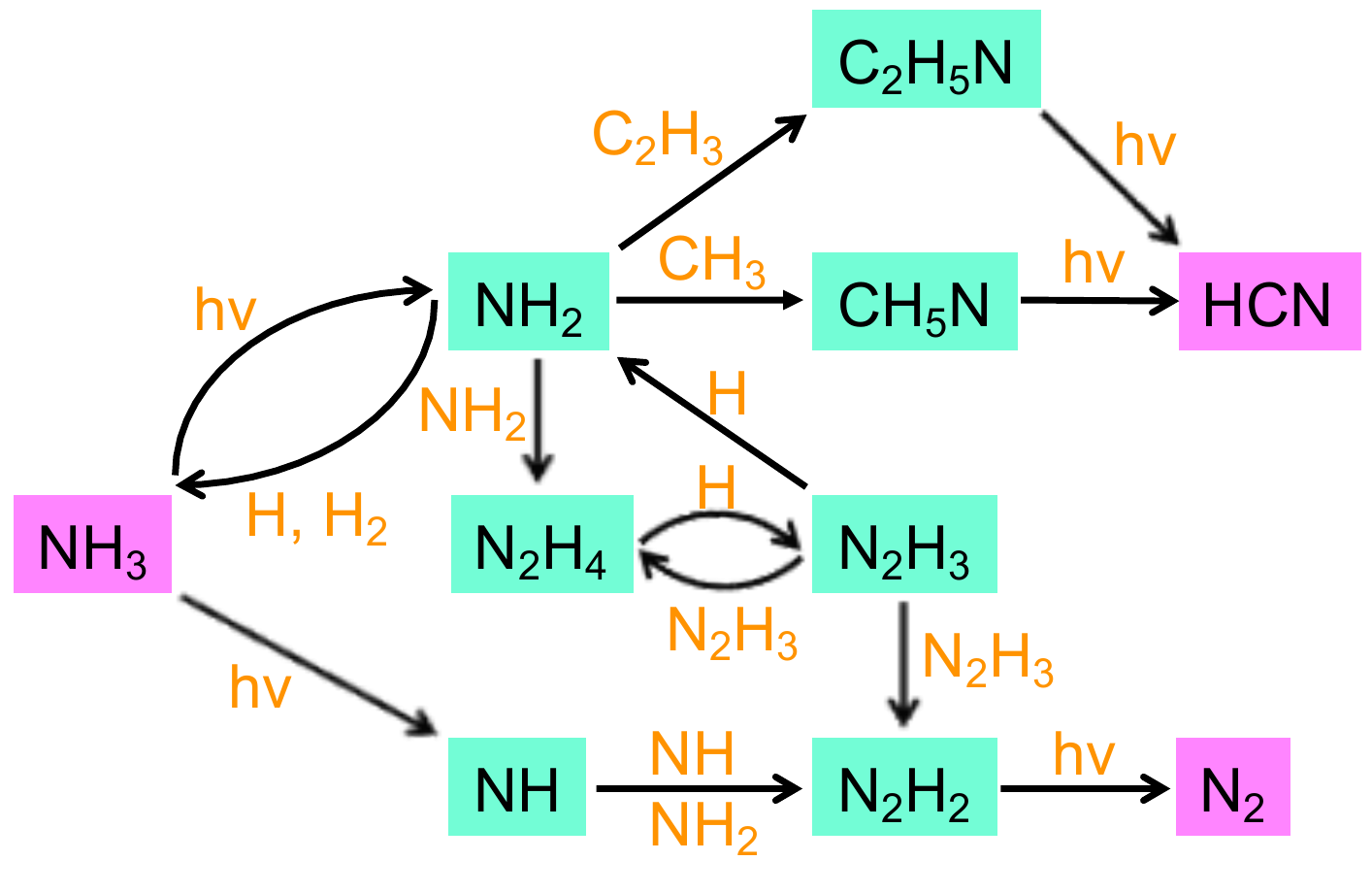}
	\caption{Chemical network from the photodissociation of \ce{NH3} in temperate and \ce{H2}-dominated atmospheres. Not all reactions are shown, and the importance of the shown reactions changes from case to case. In the presence of \ce{CH4}, \ce{HCN} is one of the main photochemical products.}
	\label{fig:nh3}
	\end{figure}
	
Figure~\ref{fig:nh3} shows the chemical network that starts with the photodissociation of \ce{NH3} and ends with the formation of \ce{N2} and \ce{HCN} as the main photochemical products. The key condition for the synthesis of \ce{HCN} is the photodissociation of \ce{NH3} in the presence of \ce{CH4} and at a temperature $>\sim200$ K. This condition allows \ce{CH3}, one of the ingredients for the synthesis of \ce{HCN}, to be produced locally by the reaction between \ce{CH4} and \ce{H}, and this \ce{H} is produced by the photodissociation of \ce{NH3} itself. We describe the details as follows.
	
The photodissociation of \ce{NH3} mainly produces \ce{NH2},
\reaction{NH3 ->[h\nu] NH2 + H, \label{r:NH3}}
and some of the \ce{NH2} produced is returned to \ce{NH3} via
\reaction{NH2 + H2 -> NH3 + H, \label{r:NH2+H2}}
and
\reaction{NH2 + H ->[M] NH3. \label{r:NH2+H}}
Another channel of the photodissociation of \ce{NH3} is to produce \ce{NH}
\reaction{NH3 ->[h\nu] NH + H2. \label{r:NH3_2}}
The \ce{NH2} channel requires photons more energetic than 230 nm and the \ce{NH} channel requires photons more energetic than 165 nm. Therefore, the photons that produce \ce{NH} is more easily shielded by \ce{H2O} and \ce{CH4}.
For the three planets modeled, the \ce{NH} channel is important in K2-18~b and Kepler-167~e, but not in PH2~b. This is because the photodissociation of \ce{NH3} occurs at higher pressures in PH2~b and is object to the shielding effect of both \ce{H2O} and \ce{CH4}. The \ce{NH} channel mostly leads to \ce{N2} (Figure \ref{fig:nh3}). 
	
The \ce{NH2} that is not recombined to form \ce{NH3} can undergo
\reaction{NH2 + NH2 -> N2H4, \label{r:NH2+NH2}}
and the \ce{N2H4} produced (if not condensed out) can then become \ce{N2H3}. \ce{N2H3} can react with itself to form \ce{N2H2}, whose photodissociation produces \ce{N2}, or with H to return to \ce{NH2} (Figure~\ref{fig:nh3}). The other loss of \ce{NH2} is to react with \ce{CH3},
\reaction{NH2 + CH3 -> CH5N, \label{r:NH2+CH3}}
followed by photodissociation to form \ce{HCN},
\reaction{CH5N ->[h\nu] HCN + 2H2. \label{r:CH5N}}
Reaction (\ref{r:NH2+CH3}) is the critical step in this \ce{HCN} formation mechanism, and it requires the \ce{CH3} radical to be available. The \ce{CH3} in Reaction (\ref{r:NH2+CH3}) is mainly produced by
\reaction{H + CH4 -> H2 + CH3. \label{r:H+CH4}}
Note that the photodissociation of \ce{CH4}, which also produces \ce{CH3}, does not contribute significantly to the source of \ce{CH3} in Reaction (\ref{r:NH2+CH3}) because the photodissociations of \ce{CH4} and \ce{NH3} typically occur at very different pressures. Another formational path of \ce{HCN} is through
\reaction{NH2 + C2H3 -> C2H5N,
\label{r:NH2+C2H3}}
followed by photodissociation
\reaction{C2H5N ->[h\nu] HCN + CH3 + H.
\label{r:C2H5N}}
The \ce{C2H3} in Reaction (\ref{r:NH2+C2H3}) is mainly produced by
\reaction{H + C2H2 ->[M] C2H3, \label{r:H+C2H2}}
and \ce{C2H2} is ultimately produced by the photodissociation of \ce{CH4} and then transported to the pressure of the photodissociation of \ce{NH3}. The \ce{HCN} produced in Reactions (\ref{r:CH5N} and \ref{r:C2H5N}) is photodissociated to form \ce{CN} but \ce{CN} quickly reacts with \ce{H2} and \ce{C2H2} to return to \ce{HCN}. Thus, \ce{HCN} does not have significant net chemical loss and is transported together with \ce{N2} through the lower boundary. 

The \ce{NH3}-\ce{CH4} coupling (Reactions~\ref{r:NH2+CH3}--\ref{r:H+CH4}) dominates the formation of \ce{HCN} over the \ce{NH3}-\ce{C2H2} coupling (Reactions~\ref{r:NH2+C2H3}--\ref{r:H+C2H2}) in temperate \ce{H2}-dominated atmospheres by several orders of magnitude. This is because the mixing ratio of \ce{C2H2} at the the pressure of \ce{NH3} photodissociation is typically very small on temperate planets like K2-18~b and PH2~b (Figures~\ref{fig:k218b_1} and \ref{fig:ph2b_1}). On colder planets like Kepler-167~e, more \ce{C2H2} is available and the \ce{NH3}-\ce{C2H2} coupling can contribute 1--10\% of the \ce{HCN} formation, consistent with the results for Jupiter \citep{moses2010abundance}. We also note that past models of warm and hot \ce{H2}-rich exoplanets suggested different reactions to represent the \ce{NH3}-\ce{CH4} coupling, including \ce{NH + CH3} \citep{line2011thermochemical} and \ce{N + CH3} \citep{kawashima2018theoretical,hobbs2019chemical}; in our models the contribution from \ce{N + CH3 -> HCN + H2} contributes to the formation of \ce{HCN} less than Reactions~(\ref{r:NH2+CH3}--\ref{r:H+CH4}) by $>3$ orders of magnitude.
	
The efficacy of the \ce{NH2} path to produce \ce{N2} and \ce{HCN} and the branching between \ce{N2} and \ce{HCN} depend on the abundance of \ce{H} and the temperature. Reaction (\ref{r:H+CH4}) has an activation energy of 33.60 kJ/mol \citep{baulch1992evaluated} and does not occur at very low temperatures. At the pressure of \ce{NH3} photodissociation, the temperature is 220 -- 240 K in K2-18~b and PH2~b, 120 -- 130 K in Kepler-167~e, and $\sim110$ K in Jupiter. This makes Reaction (\ref{r:H+CH4}) faster by six orders of magnitude in K2-18~b and PH2~b than in Kepler-167~e or Jupiter, eventually leading to an efficient \ce{HCN} production and a high abundance in the middle atmosphere. This is why the \ce{HCN} production mechanism (Reactions \ref{r:NH2+CH3}--\ref{r:H+CH4}) does not operate efficiently in giant planets in the Solar System but can build up \ce{HCN} in warmer exoplanetary atmospheres.
	
The abundance of \ce{H} is another important control. From Figure \ref{fig:nh3}, we can see that a higher abundance of \ce{H} would enhance the recycling from \ce{N2H4} to \ce{NH2}, produce more \ce{NH3} to react with \ce{NH2}, and help the return of \ce{NH2} to \ce{NH3}. In other words, a higher abundance of \ce{H} would reduce the overall efficacy of the \ce{NH2} path but favor the branch that leads to \ce{HCN}. At the pressure of \ce{NH3} photodissociation, the main source of \ce{H} is the combination of Reactions (\ref{r:NH3} and \ref{r:NH2+H2}), whose net result is the dissociation of \ce{H2} but not \ce{NH3}. The sink of \ce{H} is mainly Reaction (\ref{r:NH2+H}) and the direct recombination \ce{H + H ->[M] H2}. In high-metallicity atmospheres, another sink of \ce{H} is Reaction (\ref{r:NH2+NH2}) followed by
\reaction{N2H4 + H -> N2H3 + H2, \label{r:N2H4+H}}
and
\reaction{N2H3 + H -> 2NH2. \label{r:N2H3+H}}
The net result of Reactions (\ref{r:NH2+NH2}, \ref{r:N2H4+H}, and \ref{r:N2H3+H}) is \ce{H + H -> H2}. Therefore, the chemical network that starts with the photodissociation of \ce{NH3} is both a source and a sink of \ce{H}, which feedback to determine the outcome of the network in a non-linear way. For example, the \ce{NH2} channel is a minor pathway to form \ce{N2} in the solar or 10$\times$solar abundance atmosphere of K2-18~b but it becomes an important pathway in the $100\times$ solar atmosphere.
	
The abundance of \ce{H} at the pressure of \ce{NH3} photodissociation also explains the different sensitivity of the \ce{HCN} mixing ratio on the inclusion of deep-tropospheric source of \ce{CO}/\ce{CO} and partial depletion of \ce{NH3}. For K2-18~b, the reduction in the \ce{HCN} mixing ratio is small or proportional to the reduction in the input \ce{NH3} abundance, but more reduction in the \ce{HCN} mixing ratio is found for PH2~b (Figures~\ref{fig:k218b_1} and \ref{fig:ph2b_1}). This is because the photodissociation of \ce{NH3} occurs at higher pressures in PH2~b than in K2-18~b. When abundant \ce{CO} exists, the reactions \ce{CO + H ->[M] HCO} and \ce{HCO + H -> CO + H2} efficiently remove \ce{H}. Note that the first reaction in this cycle is three-body and only significant at sufficiently high pressures. This sink of \ce{H} results in the reduction of \ce{CH3} production (Reaction \ref{r:H+CH4}) and thus disfavors the branch in the \ce{NH2} path that leads to \ce{HCN}.
	
To summarize, the numerical models and the \ce{HCN} formation mechanism presented here indicate that \ce{HCN} and \ce{N2} are generally the expected outcomes of the photodissociation of \ce{NH3} in gaseous exoplanets that receive stellar irradiance of approximately Earth's, regardless of the stellar type.

\subsection{Photolysis of CH$_4$ Together with H$_2$O} \label{sec:ch4}
	
	\begin{figure}
	\centering
	\includegraphics[width=0.45\textwidth]{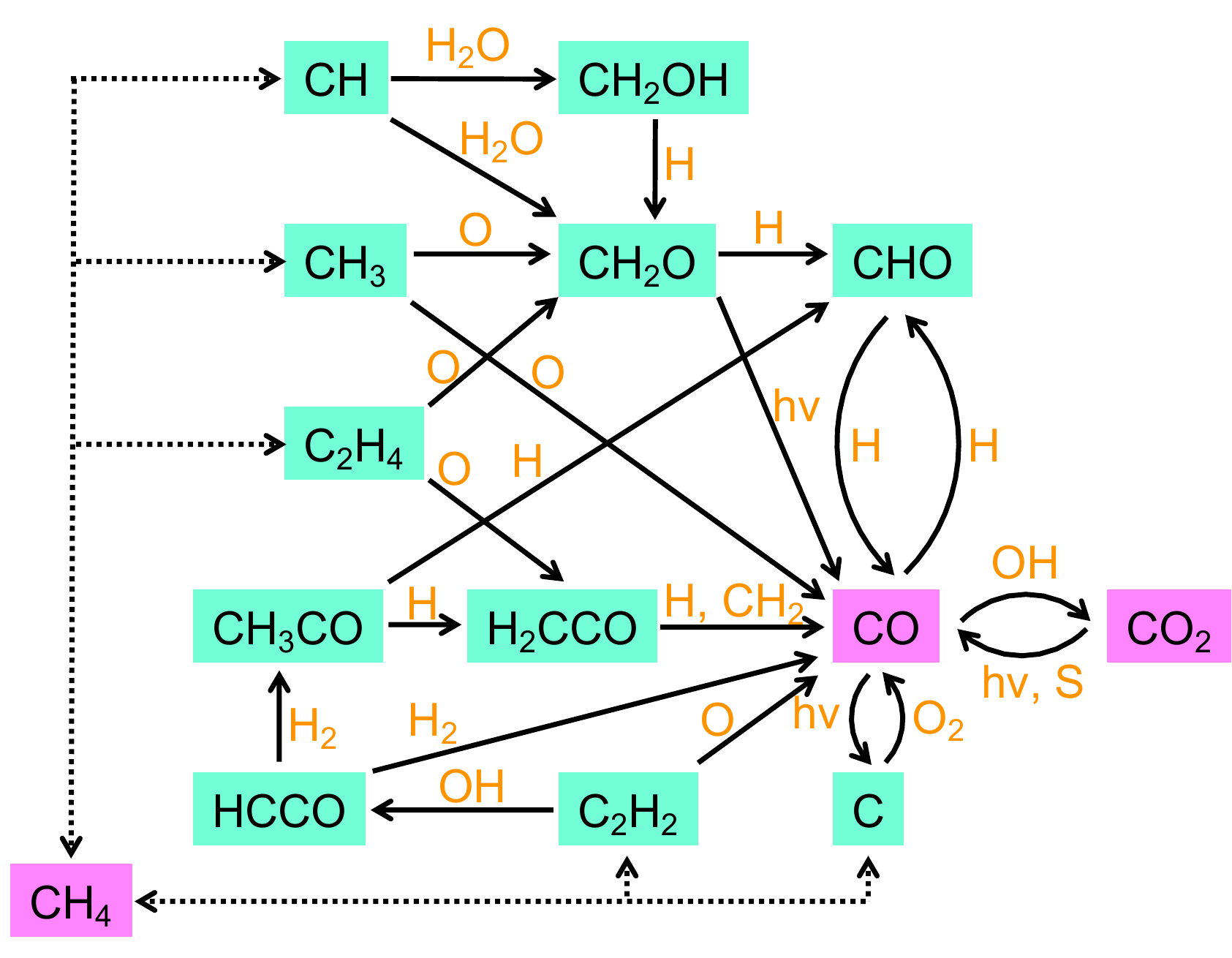}
	\caption{Chemical network from the photodissociations of \ce{CH4} and \ce{H2O} in temperate/cold and \ce{H2}-dominated atmospheres. Not all reactions are shown, and the importance of the shown reactions changes from case to case. The dashed links represent the chemical network of \ce{CH4} and hydrocarbons in cold and \ce{H2}-dominated atmospheres, as standard in the literature \citep[e.g.,][]{gladstone1996hydrocarbon,moses2005photochemistry}.	The photodissociation of \ce{H2O} provides oxidizing radicals such as \ce{OH}. When \ce{CH4} is photodissociated together with \ce{H2O}, \ce{CO} and \ce{CO2} can be formed in addition to hydrocarbons.}
	\label{fig:ch4}
	\end{figure}
	
The formation of \ce{CO} and \ce{CO2} as the most abundant photochemical gases of carbon on K2-18~b and PH2~b is another significant finding of our numerical models. The photodissociation of \ce{CH4} in colder \ce{H2}-dominated atmospheres -- such as the giant planets' atmospheres in the Solar System -- produces hydrocarbons such as \ce{C2H6} and \ce{C2H2} but not oxygenated species \citep[e.g.,][]{gladstone1996hydrocarbon,moses2005photochemistry}. This is because \ce{H2O} condenses out and is almost completely removed from the above-cloud atmosphere (such as in Kepler-167~e, Figure \ref{fig:kepler167e_1}). External sources such as comets and interplanetary dust can supply oxygen to the upper atmospheres of Jupiter and the other giant planets \citep[e.g.,][]{moses2005photochemistry,dobrijevic20201d}, but we do not include this source in the present study. For warmer planets, however, \ce{H2O} is only moderately depleted by condensation. The above-cloud water is photodissociated at approximately the same pressure as \ce{CH4} (Figures~\ref{fig:k218b_1} and \ref{fig:ph2b_1}). The photodissociations of \ce{CH4} and \ce{H2O} together in \ce{H2}-dominated atmospheres produces a chemical network beyond hydrocarbons (Figure~\ref{fig:ch4}) and eventually lead to the formation of \ce{CO} and \ce{CO2}. Even warmer atmospheres (e.g., the atmosphere of GJ~1214~b with an effective temperature of 500 -- 600 K) may also have \ce{CO} and \ce{CO2} as the most abundant photochemical gases of carbon \citep[e.g.,][]{kempton2011atmospheric,kawashima2018theoretical}.
	
The photodissociation of \ce{CH4} and the subsequent chemical reactions produce a wealth of hydrocarbons and radicals, and many of them (e.g., \ce{C}, \ce{CH}, \ce{CH3}, \ce{C2H2}, and \ce{C2H4}) lead to chemical pathways that form \ce{CO} (Figure~\ref{fig:ch4}). Between K2-18~b and PH2~b and among the modeled metallicities, we do not see a monotonic trend regarding the relative contribution of these \ce{CO} forming pathways, probably due to many chemical cycles and feedback in hydrocarbon photochemistry. \ce{CO} is converted to \ce{CO2} by the reaction with OH:
\reaction{CO + OH -> CO2 + H. \label{r:CO+OH}}
Reaction (\ref{r:CO+OH}) is the dominant source of \ce{CO2} in all models, and the only significant chemical loss of \ce{CO2} is to form \ce{CO} via either photodissociation or the reaction with elemental sulfur when available (Figure~\ref{fig:ch4}). The \ce{CO2} that is not returned to \ce{CO} is then transported through the lower boundary.
	
What are the sources of \ce{OH}, \ce{O}, and \ce{H} that power the chemical pathways shown in Figure \ref{fig:ch4}? At the pressure of \ce{CH4} and \ce{H2O} photodissociation, the source of \ce{OH} is the photodissociation of water,
\reaction{H2O ->[h\nu] H + OH,\label{r:H2O}}
and the main sink is the reaction with \ce{H2},
\reaction{OH + H2 -> H2O + H.\label{r:OH+H2}}
Reaction (\ref{r:OH+H2}) is the main sink of \ce{OH} in all models, which means that the use of \ce{OH} in the chemical pathways shown in Figure~\ref{fig:ch4} does not usually become the dominant sink of \ce{OH}. Reactions (\ref{r:H2O} and \ref{r:OH+H2}) together is equivalent to the net dissociation of \ce{H2}, which overtakes the photodissociation of \ce{CH4} and subsequent hydrocarbon reactions as the dominant source of \ce{H} in temperate atmospheres. Lastly, the main source of \ce{O} is the photodissociation of \ce{CO} and \ce{CO2}, which eventually traces to \ce{OH} and the photodissociation of water.

At this point we can explain the ratio between \ce{CO2} and \ce{CO} in the middle atmosphere, which is $\sim1$ on K2-18~b and $\sim10$ on PH2~b (Figures~\ref{fig:k218b_1} and \ref{fig:ph2b_1}). Because Reaction (\ref{r:CO+OH}) is the main source of \ce{CO2} and photodissociation is the main sink, the number density of \ce{CO2} is $\sim k_{\ref{r:CO+OH}}[\ce{CO}][\ce{OH}]/J_{\ce{CO2}}$, where $k$ is the reaction rate constant and $[]$ means the number density of a molecule. Because Reaction (\ref{r:H2O}) is the main source of \ce{OH}, the number density of \ce{OH} is $\propto J_{\ce{H2O}}[\ce{H2O}]$. Therefore, the ratio between \ce{CO2} and \ce{CO} is $\propto J_{\ce{H2O}}[\ce{H2O}]/J_{\ce{CO2}}$. For any given metallicity, the abundance of \ce{H2O} in the middle atmosphere of PH2~b is $3\sim5$-fold greater than that in K2-18~b because PH2~b is slightly warmer (Figures~\ref{fig:k218b_1} and \ref{fig:ph2b_1}). And, $J_{\ce{H2O}}$ at the top of the atmosphere on PH2~b is approximately twice that on K2-18~b, while $J_{\ce{CO2}}$ is similar between the two planets (Figure~\ref{fig:star}). Together, this causes the $[\ce{CO2}]/[\ce{CO}]$ ratio to be greater in the atmosphere of PH2~b than in K2-18~b by $\sim10$ folds.

This trend to maintain the $[\ce{CO2}]/[\ce{CO}]$ ratio also controls how the atmosphere reacts to a deep-tropospheric source of \ce{CO} and \ce{CO2} that is applied as input at the lower boundary. The input \ce{CO2} is always less than \ce{CO} by one or more orders of magnitude (Figures~\ref{fig:quench_k2_18}-\ref{fig:quench_kepler167e}). On PH2~b, photochemical processes convert \ce{CO} into \ce{CO2} in the middle atmosphere ($\sim10^2$ Pa), and cause the steady-state mixing ratio of \ce{CO2} to be greater than that of \ce{CO}. This conversion even becomes a significant sink of \ce{H2O} and causes \ce{H2O} to be depleted in the middle atmosphere in the $100\times$solar metallicity case (Figure~\ref{fig:ph2b_1}). The \ce{CO} to \ce{CO2} conversion is not so strong in the atmosphere of K2-18~b or Kepler-167~e, and their mixing ratios in the middle atmosphere are largely the input values at the lower boundary (Figure~\ref{fig:k218b_1}).
	
Finally, let us turn to the impact of \ce{H2O} and \ce{NH3} photodissociation onto the hydrocarbon chemistry. Compared with Kepler-167~e, the mixing ratio of \ce{C2H6} -- the dominant, supposedly long-lived hydrocarbon -- in K2-18~b and PH2~b is smaller and sometimes features an additional peak near the cloud deck (Figures~\ref{fig:k218b_1}-\ref{fig:kepler167e_1}). Particularly, the atmospheres of K2-18~b and PH2~b have a strong sink of \ce{C2H6} at $\sim1-10$ Pa, while the atmosphere of Kepler-167~e does not. This sink is ultimately because of the high abundance of H produced by the photodissociation of \ce{H2O} (Reactions~\ref{r:H2O} and \ref{r:OH+H2}). The detailed reaction path involves the formation of \ce{C2H5} from \ce{C2H6} (by direct reaction with \ce{H} or photodissociation to form \ce{C2H4} followed by \ce{H} addition), and then \ce{C2H5 + H -> 2CH3}. Because of the abundance of H, \ce{CH3} mostly combines with \ce{H} to form \ce{CH4}, rather than recombines to form \ce{C2H6}. It is well known that the abundance of hydrocarbons is fundamentally controlled by the relative strength between \ce{H + CH3 ->[M] CH4} and \ce{CH3 + CH3 ->[M] C2H6} \citep[e.g.][]{gladstone1996hydrocarbon,moses2005photochemistry}. Here we find that the added \ce{H} from \ce{H2O} photodissociation results in a net sink for \ce{C2H6} in K2-18~b and PH-2~b at $\sim1-10$ Pa and limits the abundance of hydrocarbons in their atmospheres. This sink does not exist in the atmosphere of Kepler-167 e, because little \ce{H2O} photodissociation occurs in its atmosphere. Additionally, near the cloud deck, the temperature is warmer, and Reaction (\ref{r:H+CH4}) that uses \ce{H} from the photodissociation of \ce{NH3} provides an additional source of \ce{CH3}, and some of the \ce{CH3} becomes \ce{C2H6} and thus its peak near the cloud deck. The formation of hydrocarbons is thus strongly impacted by the water and nitrogen photochemistry.
	
\subsection{Spectral Features of H$_2$O, CH$_4$, NH$_3$, and Photochemical Gases} \label{sec:spec}
	
\subsubsection{Transmission spectra} \label{sec:transit}
	
	\begin{figure*}
	\centering
	\includegraphics[width=0.75\textwidth]{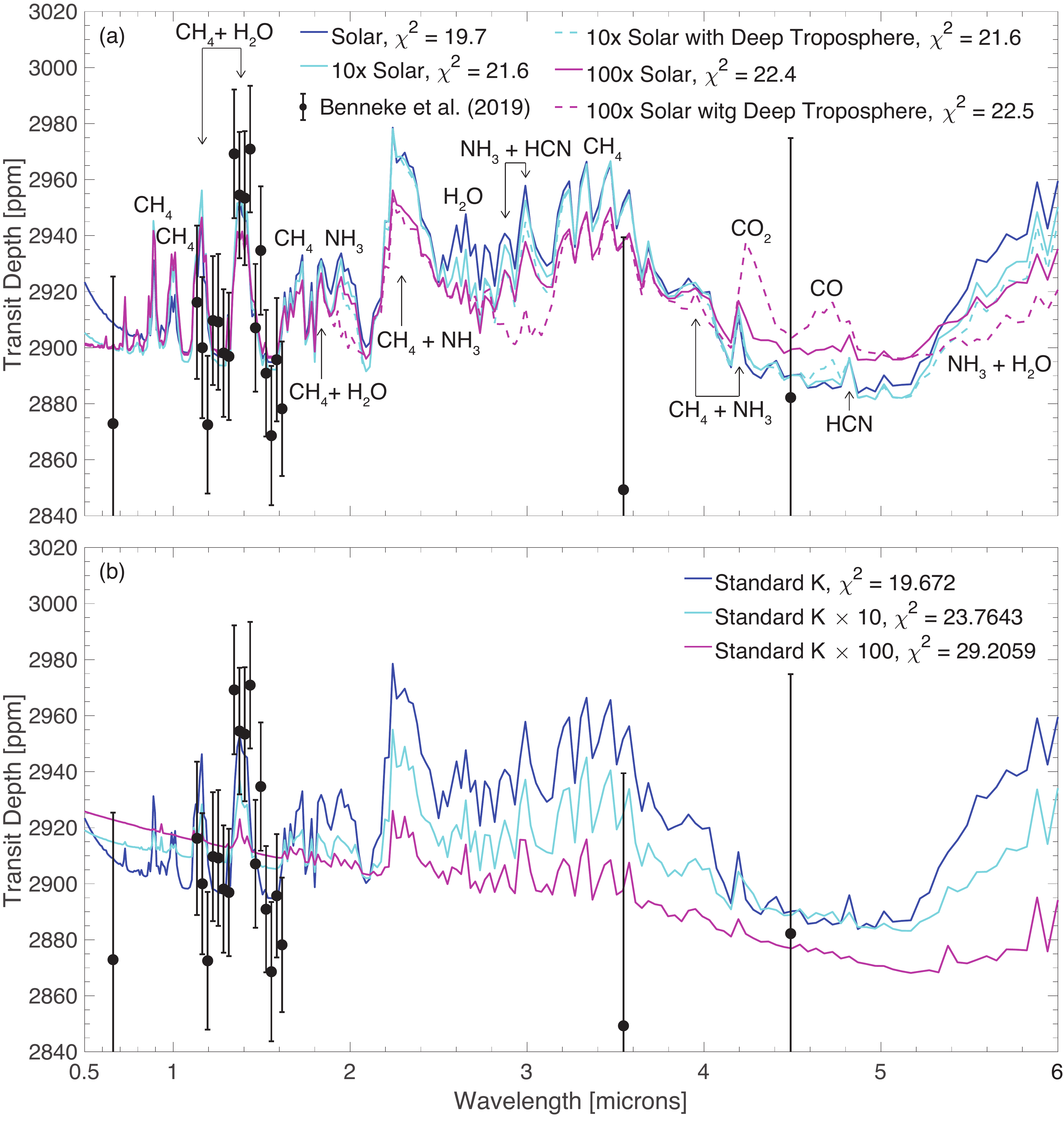}
	\caption{Modeled transmission spectra of the temperate sub-Neptune K2-18~b for varied metallicities (a) and varied eddy diffusion coefficients at the solar metallicity (b). The dashed lines show model spectra with deep-tropospheric source of \ce{CO}, \ce{CO2}, and \ce{N2} and sink of \ce{NH3}. All models with the standard eddy diffusion coefficient fit the observed transit depths. The equilibrium gases (\ce{CH4}, \ce{H2O}, and \ce{NH3}) and the photochemical gas \ce{HCN} are detectable in the wavelength range of 0.5 -- 5.0 $\mu$m. The $100\times$solar metallicity atmosphere with deep-tropospheric source and sink can have detectable features of \ce{CO2} and \ce{CO}.}
	\label{fig:k218b_spec}
	\end{figure*}
	
	\begin{figure*}
	\centering
	\includegraphics[width=0.75\textwidth]{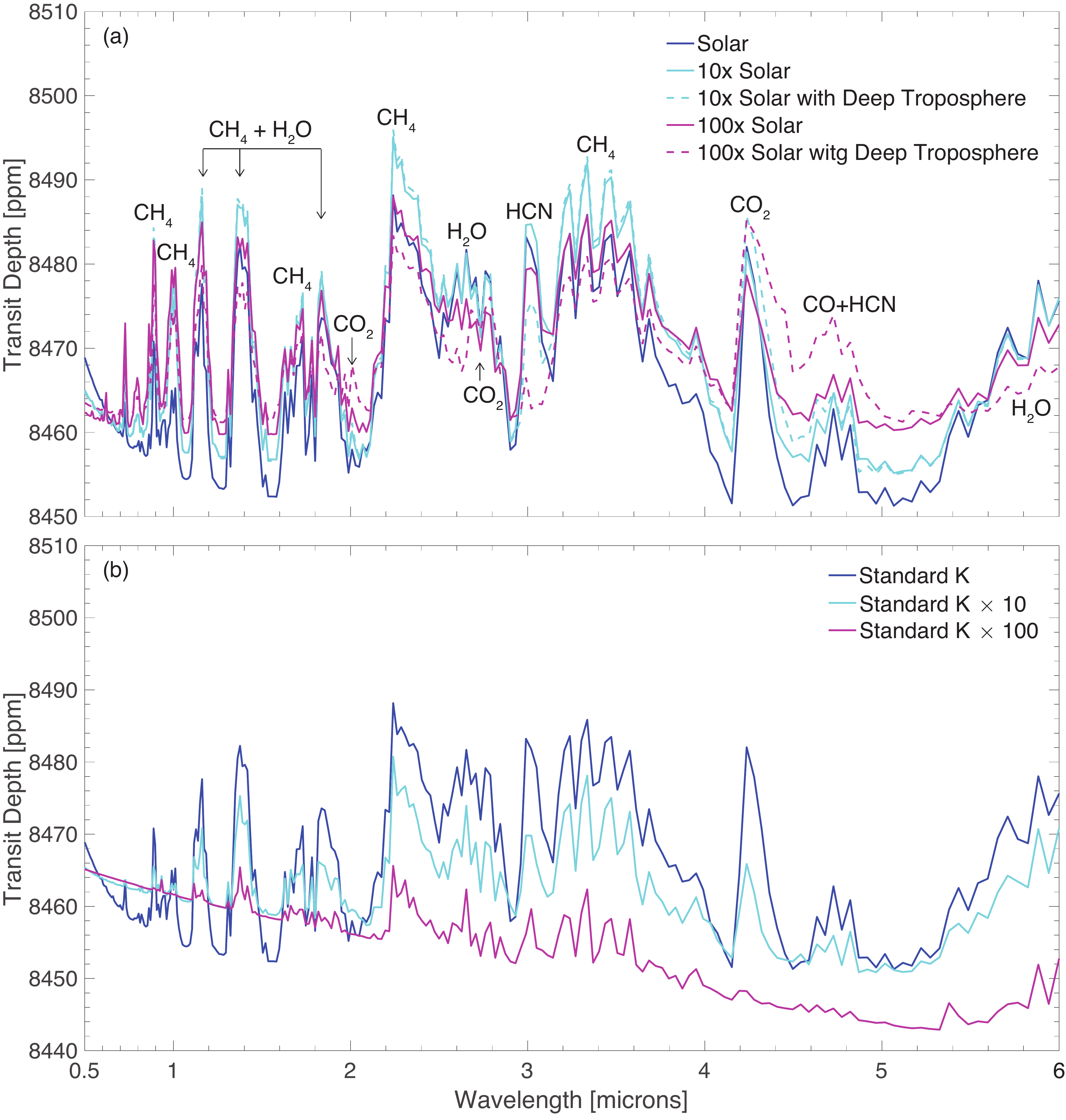}
	\caption{Modeled transmission spectra of the temperate gas giant PH2~b for varied metallicities (a) and varied eddy diffusion coefficients at the solar metallicity (b). The dashed lines show model spectra with deep-tropospheric source of \ce{CO}, \ce{CO2}, and \ce{N2} and sink of \ce{NH3}. Several equilibrium gases (\ce{CH4} and \ce{H2O}) and photochemical gases (\ce{HCN}, \ce{CO2}, and \ce{CO}) are detectable in the wavelength range of 0.5 -- 5.0 $\mu$m.}
	\label{fig:ph2b_spec}
	\end{figure*}
	
	\begin{figure*}
	\centering
	\includegraphics[width=0.75\textwidth]{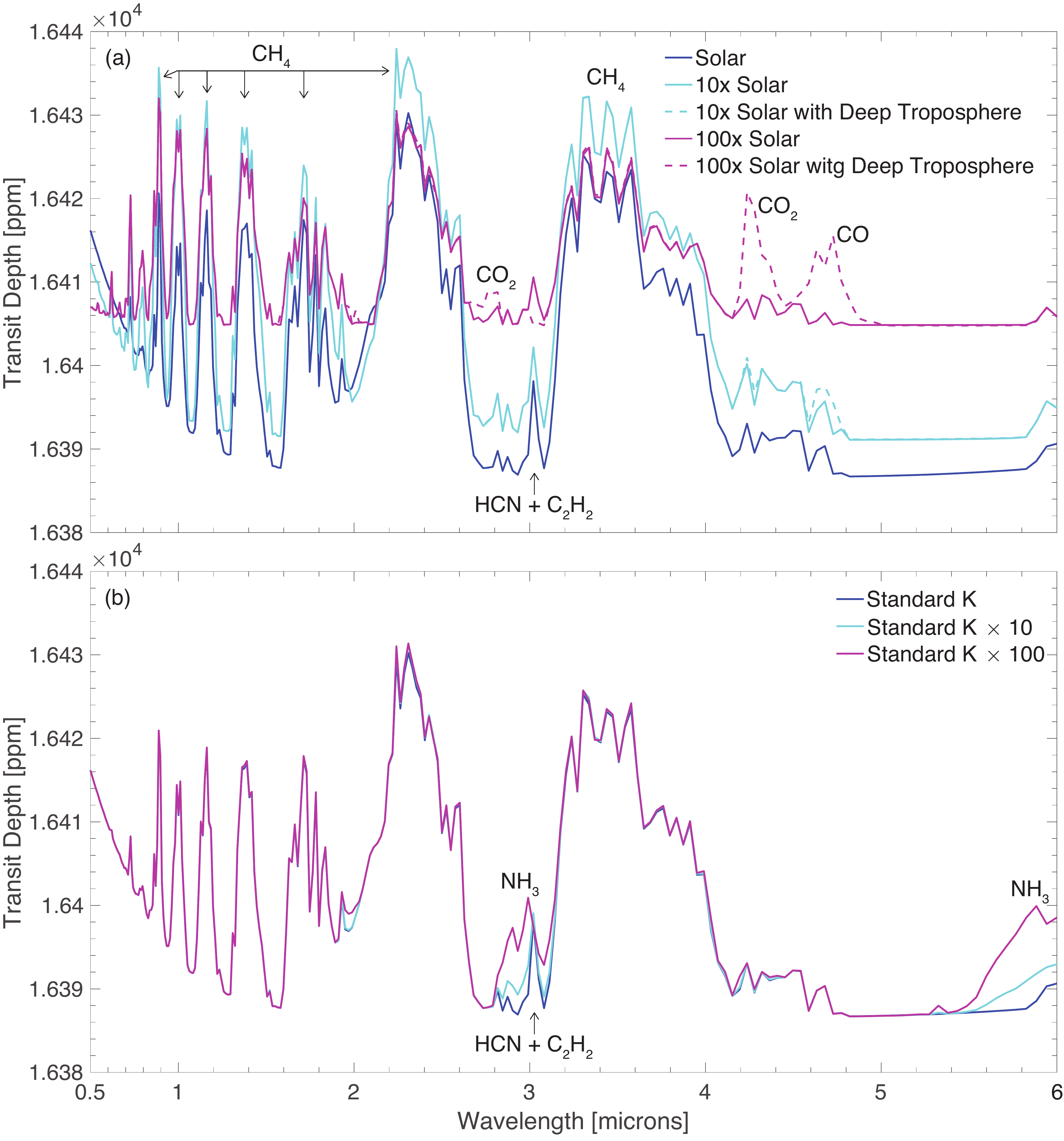}
	\caption{Modeled transmission spectra of the cold gas giant Kepler-167~e for varied metallicities (a) and varied eddy diffusion coefficients at the solar metallicity (b). The dashed lines show model spectra with deep-tropospheric source of \ce{CO}, \ce{CO2}, and \ce{N2} and sink of \ce{NH3}. With the standard eddy diffusion coefficient, \ce{CH4} is the only detectable equilibrium gas and the photochemical gases \ce{HCN} and \ce{C2H2} result in a small peak at $\sim3$ $\mu$m. The $100\times$solar metallicity atmosphere with deep-tropospheric source and sink can have detectable features of \ce{CO2} and \ce{CO}. Greater eddy diffusion coefficients can produce potentially detectable \ce{NH3}.}
	\label{fig:kepler167e_spec}
	\end{figure*}
	
Figures~\ref{fig:k218b_spec}-\ref{fig:kepler167e_spec} show the transmission spectra of the temperate and cold planets K2-18~b, PH2~b, and Kepler-167~e, based on the gas and sulfur haze profiles simulated by the photochemical models. These modeled spectra can be regarded as the canonical examples of a temperate (Earth-like insolation) planet irradiated by an M dwarf star (K2-18~b, and also TOI-1231~b), a temperate (Earth-like insolation) planet irradiated by a G/K star (PH2~b), and a cold ($\sim0.1\times$ Earth insolation) planet irradiated by a G/K star (Kepler-167~e). Here we focus on the wavelength range of 0.5 -- 5.0 $\mu$m, where several instruments on JWST will provide spectral capabilities \citep[e.g.,][]{beichman2014observations}. 
	
For K2-18~b, the equilibrium gases \ce{CH4}, \ce{H2O}, and \ce{NH3}, as well as the photochemical gas \ce{HCN} have potentially detectable spectral features in the visible to mid-infrared wavelengths (Figure~\ref{fig:k218b_spec}). Adding deep-tropospheric source of \ce{CO}, \ce{CO2}, and \ce{N2} and sink of \ce{NH3} does not cause a significant change of the spectrum of a $10\times$solar metallicity atmosphere. However, a $100\times$solar metallicity atmosphere with deep-tropospheric source and sink would be free of the spectral features of \ce{NH3} or \ce{HCN}, but instead have potentially detectable features of \ce{CO2} and \ce{CO}.

Strikingly, the models from $1\times$ to $100\times$ solar abundance and with the standard eddy diffusion coefficient provide good fits to the existing transit depth measurements by K2, \textit{Hubble}, and \textit{Spitzer} \citep{tsiaras2019water,benneke2019water}. The models with a 100-fold greater eddy diffusion coefficient would have the sulfur haze layer extending to $10^2$ Pa and mute the spectral features in 1.1 -- 1.7 $\mu$m, at odds with the \textit{Hubble} data. Both \ce{CH4} and \ce{H2O} contribute to the spectral modulations seen by \textit{Hubble}, which may have caused the difficulties in the identification of the gases by spectral retrieval \citep{tsiaras2019water,benneke2019water,Blain2020}.
	
\ce{HCN}, one of the most abundant photochemical gases in the middle atmosphere, is likely detectable in K2-18~b via its spectral band at $\sim3.0$ $\mu$m. The \ce{HCN} is produced from the photodissociation of \ce{NH3} in presence of \ce{CH4}. Also at $3.0$ $\mu$m are the absorption bands of \ce{NH3} and to a lesser extent \ce{C2H2}. It would be possible to disentangle these bands with a reasonably wide wavelength coverage because \ce{NH3} has multiple and more prominent bands in the mid-infrared (Figure~\ref{fig:k218b_spec}), and because \ce{C2H2} should have a minimal abundance in the middle atmosphere (Figure~\ref{fig:k218b_1}) and contribute little to the transmission spectra. 
	
The spectral bands of \ce{CO2} and \ce{CO} can be seen in the modeled spectra (in 4 -- 5 $\mu$m) of K2-18~b only when the atmosphere has super-solar metallicity and the transport from the deep troposphere is taken into account (Figure~\ref{fig:k218b_spec}). In other words, the \ce{CO} and \ce{CO2} that are produced from the photodissociation of \ce{CH4} together with \ce{H2O} would have too low mixing ratios to be detected. The photodissociation of \ce{CH4} also produces \ce{C2H6}. While \ce{C2H6} has strong bands at 3.35 and 12 $\mu$m, they would not be detectable due to its relatively low abundance and the strong \ce{CH4} and \ce{NH3} bands at the same wavelength, respectively (Figure~\ref{fig:k218b_spec}).
	
For PH2~b, prominent spectral bands of \ce{CH4}, \ce{H2O}, and the photochemical gases \ce{CO2} and \ce{HCN} can be expected (Figure~\ref{fig:ph2b_spec}). \ce{NH3} is not detectable because it is depleted by photodissociation to the cloud deck (Figure~\ref{fig:ph2b_1}). Even though its pressure of photochemical depletion can be reduced to $\sim10^2$ Pa for a large eddy diffusion coefficient, the sulfur haze in that case would mute spectral features that are generated from approximately the same pressure levels (Figure~\ref{fig:k}) and thus cause \ce{NH3} to be undetectable. \ce{HCN}, \ce{CO2}, and \ce{CO} are the most abundant photochemical gases (Figure~\ref{fig:ph2b_1}); but the \ce{CO} bands are intrinsically weaker and so \ce{CO2} and \ce{HCN} are the detectable photochemical gases via their spectral bands at 4.2 and 3.0 $\mu$m, respectively. Similar to K2-18~b, adding deep-tropospheric source of \ce{CO}, \ce{CO2}, and \ce{N2} and sink of \ce{NH3} does not cause a significant change of the spectrum of a $10\times$solar metallicity atmosphere. However, a $100\times$solar metellicity atmosphere with deep-tropospheric source and sink would not have the spectral features of \ce{H2O} or \ce{HCN} and have more prominent features of \ce{CO2} and \ce{CO}, as predicted by the photochemical model (Figure~\ref{fig:ph2b_1}).
	
Lastly for the cold planet Kepler-167~e, the transmission spectra will be dominated by the absorption bands of \ce{CH4} (Figure \ref{fig:kepler167e_spec}), as \ce{H2O} is completely removed by condensation and \ce{NH3} by condensation and photodissociation. For a large eddy diffusion coefficient, the pressure of photochemical depletion of \ce{NH3} can be reduced to $\sim10^3$ Pa (Figure~\ref{fig:k}) and this can produce a spectral band of \ce{NH3} at $\sim3.0$ $\mu$m. Thus, a search for this absorption band in the transmission spectra may constrain the eddy diffusion coefficient, although to distinguish it with a small peak due to the combined absorption of the photochemical gases \ce{HCN} and \ce{C2H2} (Figure~\ref{fig:kepler167e_spec}) may involve quantification through photochemical models. The main photochemical gas in this cold atmosphere \ce{C2H6} has spectral bands at 3.35 and 12 $\mu$m. The 3.35-$\mu$m band is buried by a strong \ce{CH4} band, and while not shown in Figure~\ref{fig:kepler167e_spec}, the 12-$\mu$m band might be detectable given appropriate instrumentation with the spectral capability in the corresponding wavelength range. Finally, the deep-troposphere-sourced \ce{CO2} and \ce{CO} in a $100\times$solar metallicity atmosphere may produce detectable spectral features in 4 -- 5 $\mu$m.
	
To summarize, transmission spectroscopy from the visible to mid-infrared wavelengths can provide the sensitivity to detect the equilibrium gases \ce{CH4} and \ce{H2O}, and the photochemical gases \ce{HCN}, and in some cases \ce{CO2} in temperate/cold and \ce{H2}-rich exoplanets. We do not expect \ce{C2H6} to be detectable. \ce{NH3} would be detectable on temperate planets around M dwarf stars but not detectable on temperate planets around G/K stars. The deep-tropospheric source and sink can have a major impact only on the transmission spectrum of a $100\times$solar metallicity atmosphere, where typically the features of \ce{NH3} and \ce{HCN} would be reduced and those of \ce{CO2} and \ce{CO} would be amplified. The detection and non-detection of these gases will thus test the photochemical model and improve our understanding of the photochemical mechanisms as well as tropospheric transport in temperate/cold and \ce{H2}-dominated atmospheres.
	
\subsubsection{Spectra of the reflected starlight} \label{sec:reflect}
	
	\begin{figure*}
	\centering
	\includegraphics[width=0.95\textwidth]{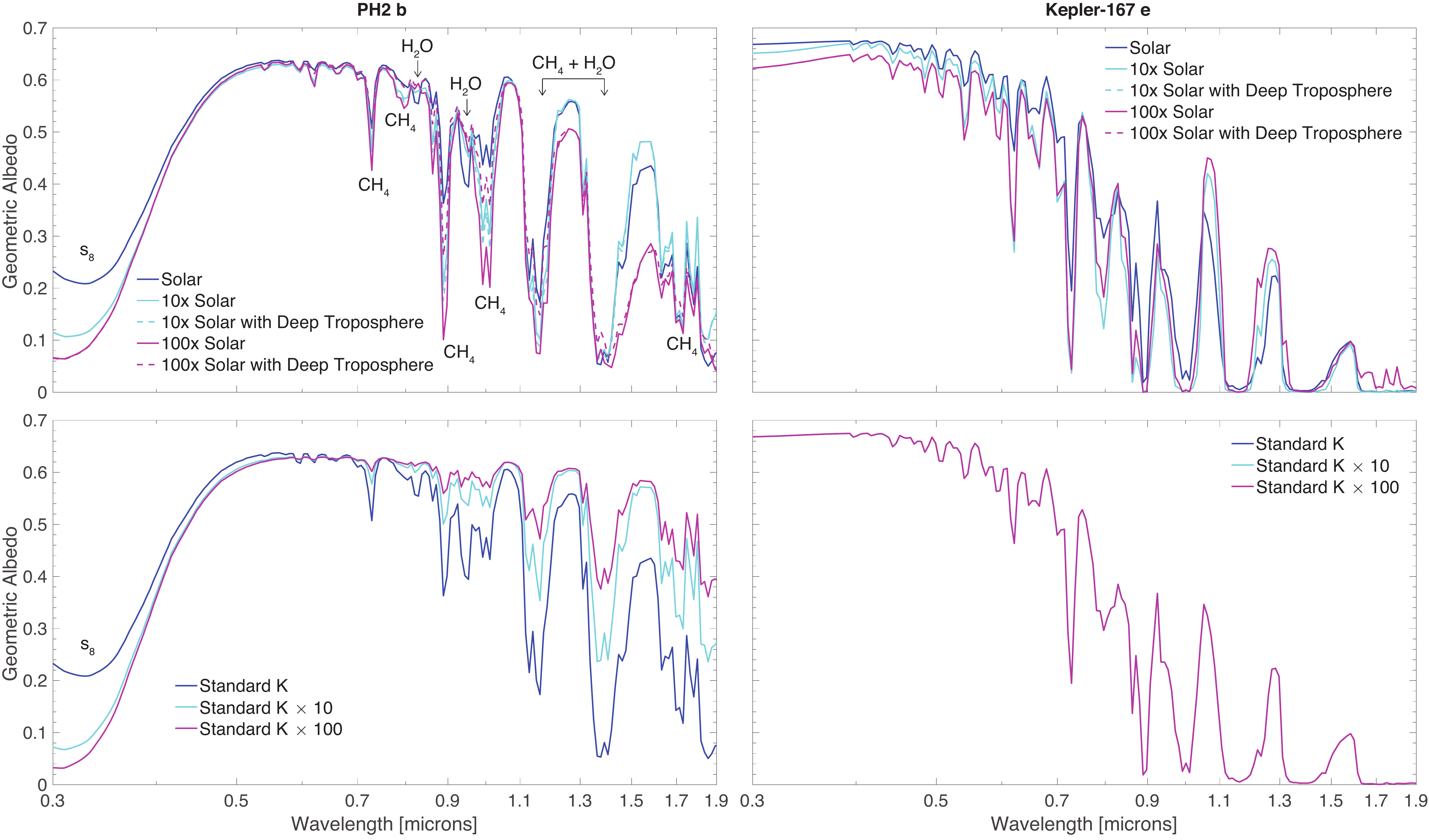}
	\caption{Modeled geometric albedo spectra of the temperate gas giants PH2~b and the cold gas giant Kepler-167~e for varied metallicities and varied eddy diffusion coefficients at the solar metallicity. The dashed lines show model spectra with deep-tropospheric source of \ce{CO}, \ce{CO2}, and \ce{N2} and sink of \ce{NH3}. All absorption features in Kepler-167~e's spectra are due to \ce{CH4}. Both \ce{H2O} and \ce{CH4} can be detectable in the reflected starlight of PH2~b and only \ce{CH4} can be detectable in Kepler-167~e.}
	\label{fig:al}
	\end{figure*}
	
The temperate and cold planets around G/K stars are widely separated from their host stars and may thus also be characterized in the reflected starlight by direct imaging. Figure~\ref{fig:al} shows the geometric albedo spectra of PH2~b and Kepler-167~e in the visible and near-infrared wavelengths that approximately correspond to the Roman Space Telescope's coronagraph instrument \citep{kasdin2020nancy} and its potential Starshade Rendezvous \citep{Seager2019} and the HabEx concept \citep{Gaudi2020}. While PH2~b and Kepler-167~e themselves are not potential targets for these missions, their albedo spectra broadly resemble the targets in the temperate (PH2~b) and cold (Kepler-167~e) regimes.
	
The spectral features of \ce{CH4} and \ce{H2O} can be seen in the reflected starlight of PH2~b. This ability to detect \ce{H2O} in giant planets warmer than Jupiter is consistent with \cite{MacDonald2018}. In addition to the absorption features of \ce{CH4} and \ce{H2O}, the albedo spectra of PH2~b feature the absorption of the sulfur (\ce{S8}) haze layer at wavelengths shorter than $\sim0.5$ $\mu$m. This result is consistent with the findings of \cite{gao2017sulfur}. For a greater eddy diffusion coefficient, the sulfur haze layer is higher and the spectral features of \ce{CH4} and \ce{H2O} become weaker. Interestingly, the absorption features of \ce{H2O} are the most prominent in the solar-abundance case, and they are somewhat swamped by the adjacent \ce{CH4} features at higher metallicities. This is because, as \ce{H2O} condenses out, the above-cloud mixing ratio of \ce{H2O} only slightly increases with the metallicity, while that of \ce{CH4} increases proportionally (Figure~\ref{fig:ph2b_1}). Only the absorption of \ce{CH4} can be seen in the albedo spectra of Kepler-167~e, as \ce{H2O} is depleted by condensation. On both planets, the spectral features of \ce{NH3} are not seen due to its weak absorption \citep{irwin2018analysis} and photochemical depletion to the cloud deck (Figures~\ref{fig:ph2b_1} and \ref{fig:kepler167e_1}). The deep-tropospheric source and sink has minimal impact on the albedo spectra, unless in the $100\times$solar metallicity atmosphere on PH2~b where a reduction of the \ce{CH4} features can be seen.
	
	
	
\section{Discussion} \label{sec:discussion}
	
The results and analyses presented in Section \ref{sec:result} indicate that the temperate and \ce{H2}-rich exoplanets, particularly those orbiting M dwarf stars, provide an unprecedented opportunity to characterize the photochemical mechanisms in low-temperature atmospheres. So far, these planets include K2-18~b, TOI-1231~b, and LHS-1140~b if they have \ce{H2}-dominated atmospheres. \ce{H2}-dominated atmospheres that receive stellar irradiation as Earth are not found in the Solar System, and we have shown here that this unique exoplanetary regime would result in mechanisms to form \ce{HCN} as a uniformly abundant product from the photodissociation of \ce{NH3} in presence of \ce{CH4}, as well as detectable levels of \ce{CO2} on planets around G/K stars from the photodissociation of \ce{CH4} together with \ce{H2O}. The observations of temperate and \ce{H2}-rich exoplanets thus promise to greatly expand the types of molecules detected in exoplanet atmospheres. Interestingly, \ce{HCN} is one of the most important molecules for prebiotic chemistry \citep[e.g.,][]{patel2015common}, and the exoplanet observations may constrain the photochemical pathways for its formation in primordial planetary atmospheres.
	
For K2-18~b, our model predicts that the spectral features of \ce{CH4} can have a size of $\sim80$ ppm in the transit depth, and those of \ce{H2O}, \ce{NH3}, \ce{HCN}, and \ce{CO2} (from the deep troposphere) would have a size of $30\sim60$ ppm. These quantities are substantially above the current estimate of the potential ``noise floor'' of the near-infrared instruments on JWST \citep[$<\sim10$ ppm,][]{schlawin2020jwst,schlawin2021jwst}, and are thus likely measurable. These spectral features may also be within the reach of ARIEL \citep{tinetti2018chemical,changeat2020disentangling}.
	
An an example, we have used PandExo \citep{batalha2017pandexo} to estimate the overall photometric uncertainties achieved by observing the transits of K2-18~b with the G235H and G395H gratings of the NIRSpec instrument on JWST. These two channels would cover the wavelength range of $1.7-5.2\ \mu$m and thus provide the sensitivity to the spectral features shown in Figure~\ref{fig:k218b_spec}. We find that with two visits in G235H and four visits in G395H, the overall photometric precision would be $\sim20$ ppm per spectral element at the resolution of $R=100$ in both wavelength channels, and this precision should enable the detection of \ce{CH4}, \ce{H2O}, \ce{NH3}, the photochemical gas \ce{HCN}, and possibly \ce{CO2}. If reducing the spectral resolution to $R=50$, the number of visits would be halved, but this could cause spectral ambiguity between \ce{NH3} and {HCN} because they both have absorption bands at $\sim3.0$ $\mu$m (Figure~\ref{fig:k218b_spec}). Spectral ambiguity in the transmission spectra with the resolution of $R\sim50$ or less has been recently shown with \textit{Hubble} at $1.1-1.7\ \mu$m \citep[][]{mikal2020transmission}.
	
The size of the transmission spectral features expected for temperate and cold gas giants around G/K stars, such as PH2~b and Kepler-167~e, is small but probably not prohibitive. For example, our model predicts that the spectral features of \ce{CH4} can have a size of $\sim50$ ppm in the transit depth, and those of \ce{H2O}, \ce{CO2}, and \ce{HCN} would have a size of $20\sim30$ ppm. Several visits may need to be combined to achieve the photometric precision to detect these gases. Complementary to transmission spectroscopy, future direct-imaging missions can readily detect \ce{CH4}, \ce{H2O}, and clouds \citep[e.g.,][]{damiano2020exorel}, as well as the sulfur haze produced by atmospheric photochemistry.
	
While we focus on temperate and cold planets in this paper, the photochemical mechanisms and the predictions on the gas formation and spectral features should remain applicable to the planets that are only slightly warmer than K2-18~b and PH2~b. This is because the results on these planets do not rely on the formation of water clouds. We suspect that the results should be applicable as long as the dominant O, C, N, S species in thermochemical equilibrium with \ce{H2} are \ce{H2O}, \ce{CH4}, \ce{NH3}, and \ce{H2S} and the assumptions on other atmospheric parameters (e.g., the eddy diffusion coefficient) remain broadly valid. 
	
The eddy diffusion coefficient adopted in this work corresponds to that of Jupiter \citep{conrath1984global} and features a minimum at the bottom of the stratosphere. This minimum value is also close to the eddy diffusion coefficient at the troposphere-stratosphere boundary of Earth's atmosphere \citep{massie1981stratospheric}. However, the adopted eddy diffusion coefficient at the bottom of the stratosphere is smaller than the values used in past photochemical models of warmer exoplanets \citep[e.g., GJ~1214~b and GJ~436~b,][]{kempton2011atmospheric,Moses2013,Hu2014B2014ApJ...784...63H} or the values derived from a 3D particulate tracer-transport model conditioned on hot Jupiters \citep{parmentier20133d} by several orders of magnitude. We note that Earth, the cold giant planets in the Solar System, and the modeled K2-18~b \citep{Blain2020,charney2020} all have temperature inversion and thus a true stratosphere, while atmosphere models of the warm exoplanets GJ~1214~b and GJ~436~b do not predict temperature inversion \citep[e.g.,][]{kempton2011atmospheric,Moses2013}. The lower temperature and the temperature inversion may both contribute to the lower eddy diffusion coefficient on temperate and cold exoplanets. Predictive models of the eddy diffusion coefficient in exoplanets are being developed \citep[e.g.,][]{zhang2018global1,zhang2018global2} and can be tested by future observations as shown in Figures \ref{fig:k218b_spec}-\ref{fig:kepler167e_spec}.

We have also shown in Section~\ref{sec:result} that the deep-tropospheric source of \ce{CO}, \ce{CO2}, and \ce{N2} and sink of \ce{NH3} can substantially change the composition of the observable part of the atmosphere -- and the transmission spectrum -- if the atmosphere has $100\times$solar metallicity. The main change is the reduction of \ce{NH3} and \ce{HCN} and the enhancement of \ce{CO} and \ce{CO2} in the spectrum. As such, detecting and measuring the abundance of these gases in the temperate \ce{H2}-dominated atmosphere may provide constraints on the temperature and the strength of vertical mixing in the deep troposphere \citep[e.g.,][]{fortney2020beyond}. One should note that modification of the deep-tropospheric abundance of gases by photochemical processes will be important in this endeavor: \ce{NH3} is expected to be depleted anyway and \ce{CO2} should overtake \ce{CO} as the main carbon molecule in the middle atmosphere of temperate and \ce{H2}-rich exoplanets of G/K stars.

A recently published study of atmospheric photochemistry in the atmosphere of K2-18~b \citep{yu2021identify} came to our notice during the peer-review phase of this work. The ``no-surface'' case in \cite{yu2021identify} has a comparable physical picture as the $100\times$solar metallicity case with the deep-tropospheric source and sink presented in Figure~\ref{fig:k218b_1}. A common feature is that such an atmosphere would be rich in \ce{CO} and \ce{CO2}, and the difference in the profiles of \ce{HCN} and other photochemical gases between the models may be due to the assumed profile of eddy diffusivity.
	
Lastly, we emphasize that several effects of potential importance have not been studied in this work. A more accurate pressure-temperature profile from 1D or 3D models may improve the prediction on the extent of water vapor depletion by condensation. A temperature inversion would result in higher temperatures in the upper stratosphere than what has been adopted here, and this may have an impact on the efficacy and relative importance of chemical pathways. A more accurate pressure-temperature profile and vertical mixing modeling for the deep troposphere may improve the prediction and perhaps remove the need for the endmember scenarios as presented. On planets that are expected to be tidally locked, the transmission spectra are controlled by the chemical abundance at the limb \citep[e.g.,][]{steinrueck2019effect,drummond2020implications}, and thus the horizontal transport of long-lived photochemical gases such as \ce{HCN} and \ce{CO2} may be important. Finally, we have not included hydrocarbon haze in this study, while it can form with both \ce{C2H2} and \ce{HCN} in the atmosphere \citep{kawashima2019detectable}. We hope that the present work will help motivate future studies to address these potential effects.
	
\section{Conclusion} \label{sec:conclusion}
	
We have studied the photochemical mechanisms in temperate/cold and \ce{H2}-rich exoplanets. For the \ce{H2}-rich planets (giants and mini-Neptunes) that receive stellar irradiance of approximately Earth's, we find that the main photochemical gases are \ce{HCN} and \ce{N2}. The synthesis of \ce{HCN} requires the photodissociation of \ce{NH3} in presence of \ce{CH4} at a temperature $>\sim200$ K. \ce{NH3} is dissociated near the water cloud deck and thus has a minimal mixing ratio in the middle atmosphere (10 -- $10^3$ Pa) if the planet orbits a G/K star, but \ce{NH3} can remain intact in the middle atmosphere if the planet orbits an M star. Additional photochemical gases include \ce{CO}, \ce{CO2}, \ce{C2H6}, and \ce{C2H2}. \ce{CO} and \ce{CO2} are the main photochemical gas of carbon because of the photodissociation of \ce{H2O} together with \ce{CH4}. The photodissociation of \ce{H2O} also strongly limits the abundance of photochemical hydrocarbons in the atmosphere. For the planets that receive stellar irradiance of approximately $0.1\times$ Earth's, the formation of \ce{HCN} is limited by the low temperature, \ce{CO2} or \ce{CO} is not produced due to nearly complete removal of \ce{H2O} by condensation, and the main photochemical gases are \ce{C2H6} and \ce{C2H2}.
	
The photochemical models of the temperate sub-Neptune K2-18~b assuming $1-100\times$solar abundance result in transmission spectra that fit the current measurements from K2, \ce{Hubble}, and \ce{Spitzer}. Both \ce{CH4} and \ce{H2O} contribute to the spectral modulation seen by \ce{Hubble}. Transmission spectroscopy with JWST and ARIEL will likely provide the sensitivity to detect the equilibrium gases \ce{CH4}, \ce{H2O}, and \ce{NH3}, the photochemical gas \ce{HCN}, and in some cases \ce{CO2}. \ce{C2H6} is unlikely to be detectable due to its low mixing ratio and spectral feature overwhelmed by \ce{CH4}. Transmission spectroscopy of the temperate giant planets around G/K stars will likely provide the sensitivity to detect \ce{CH4}, \ce{H2O}, and the photochemical gases \ce{HCN} and \ce{CO2}, complementing future spectroscopy in the reflected light by direct imaging. If the eddy diffusion coefficient is greater than that in Jupiter by two orders of magnitude, the sulfur haze layer would subdue the transmission spectral features -- but this situation is unlikely for K2-18~b because of the detected spectral modulation. These results are also applicable to similarly irradiated \ce{H2}-rich exoplanets, including TOI-1231~b and LHS-1140~b if they have \ce{H2}-dominated atmospheres.
	
The results here indicate that the temperate/cold and \ce{H2}-rich exoplanets, which often represent a temperature and atmospheric composition regime that is not found in the Solar System, likely have rich chemistry above clouds that leads to a potpourri of photochemical gases, some of which will build-up to the abundance detectable by transmission spectroscopy soon. The detection of atmospheric photochemical products in K2-18~b and other temperate exoplanets would expand the types of molecules detected in exoplanet atmospheres and greatly advance our understanding of the photochemical processes at works in low-temperature exoplanets. 
	
\section*{Acknowledgments}
We thank Sara Seager and the anonymous referee for helpful comments that improved the paper, Yuk Yung and Danica Adams for providing information on the Caltech/JPL KINETICS code for comparison, and Mario Damiano for advice on the PandExo simulations. This work was supported in part by the NASA Exoplanets Research Program grant \#80NM0018F0612. This research was carried out at the Jet Propulsion Laboratory, California Institute of Technology, under a contract with the National Aeronautics and Space Administration. 
	
	{	\small
		\bibliographystyle{apj}
		\bibliography{bib.bib}
	}
	
	
\end{document}